\documentclass{aa}  

\usepackage{graphicx}
\usepackage{xcolor}
\usepackage{txfonts}
\usepackage{multirow}
\usepackage{rotating}

\usepackage[hidelinks,colorlinks=true,linkcolor=red,citecolor=blue,urlcolor=violet]{hyperref}
\usepackage{natbib}

\defcitealias{Bartalesi24}{B24}
\defcitealias{Bartalesi25}{B25}
\defcitealias{Ghirardini19}{G19}
\defcitealias{angelinelli20}{A20}
\defcitealias{XRISM_a2029_2_25}{X25a}
\defcitealias{BALDI17}{B17}
\defcitealias{XRISM_simulations}{X25c}

\newcommand{\nstar}{n_{\star}}
\newcommand{\Tstar}{T_{\star}}
\newcommand{\Rstar}{R_{\star}}
\newcommand{\Rcen}{R_{\rm{CEN}}}
\newcommand{\gamIN}{\gamma'_{\rm{IN}}}
\newcommand{\gamOUT}{\gamma'_{\rm{OUT}}}
\newcommand{\gamCEN}{\gamma'_{\rm{CEN}}}

\newcommand{\proton}{m_{\rm p}}
\newcommand{\Phistar}{\Phi_{\rm{eff,\star}}}
\newcommand{\PhiCEN}{\Phi_{\rm{eff,CEN}}}
\newcommand{\Phieff}{\Phi_{\rm{eff}}}
\newcommand{\Phieffzero}{\Phi_{\rm{eff,0}}}
\newcommand{\rfive}{r_{500}}
\newcommand{\Sm}{S_{m}}
\newcommand{\zup}{z_{{\rm up},m}}
\newcommand{\ylow}{y_{{\rm low},m}}
\newcommand{\yup}{y_{{\rm up},m}}

\newcommand{\Rhat}{\hat R}
\newcommand{\phihat}{\hat \phi}
\newcommand{\Ri}{\hat R_{i}}
\newcommand{\Rplus}{\hat R_{i + 1}}

\newcommand{\kms}{\rm{km \, s^{-1}}}

\newcommand{\Mtwo}{M_{200}}
\newcommand{\rhocrit}{\rho_{\rm{crit}}}
\newcommand{\diff}{\mathrm{d}}
\newcommand{\nee}{n_{\rm{e}}}

\newcommand{\Tgas}{T_{\rm{gas}}}
\newcommand{\nH}{n_{\rm{H}}}
\newcommand{\Dang}{D_{\rm{ang}}}
\newcommand{\sigmagas}{\sigma_{\rm{gas,1D}}}
\newcommand{\sigmaturb}{\sigma_{\rm{turb,1D}}}

\newcommand{\kboltz}{k_{\rm{B}}}
\newcommand{\alphainf}{\alpha_{\rm{inf}}}
\newcommand{\alphazero}{\alpha_0}
\newcommand{\Rpeak}{R_{\rm{peak}}}
\newcommand{\upeak}{u_{\rm{peak}}}
\newcommand{\alphaturb}{\alpha_{\rm{turb}}}

\newcommand{\Normizero}{Norm_{i,0}}

\newcommand{\sigmavi}{\sigma_{{\rm v},i}}

\newcommand{\Nh}{N_{\rm{H}}}

\newcommand{\Rcore}{R_{\rm{c}}}
\newcommand{\uewm}{u_{{\rm ew},m}}

\newcommand{\Rbar}{\bar{R}}
\newcommand{\phibar}{\bar{\phi}}
\newcommand{\pturb}{P_{\rm{turb}}}
\newcommand{\Iiw}{I_{i,w}}
\newcommand{\Iizero}{I_{i,0}}

\newcommand{\RSZ}{r_{\rm{SZ}}}
\newcommand{\sigmavm}{\sigma_{{\rm v}, m}}
\newcommand{\PSZj}{P_{{\rm SZ},j}}
\newcommand{\radj}{r_{j}}
\newcommand{\radplus}{r_{j+1}}
\newcommand{\arcminute}{\rm{arcmin}}
\newcommand{\Niii}{\mathcal{N}_{i}}
\newcommand{\sigmastat}{\sigma_{\rm{stat}}}
\newcommand{\epsN}{\epsilon_{\rm{N}}}
\newcommand{\epsT}{\epsilon_{\rm{T}}}
\newcommand{\epsSZ}{\epsilon_{\rm{SZ}}}
\newcommand{\zum}{z_{{\rm u},m}}
\newcommand{\Ti}{T_{i}}
\newcommand{\Tiw}{T_{i,w}}
\newcommand{\Zi}{Z_{i}}
\newcommand{\uiw}{u_{i,w}}

\newcommand{\Normiw}{Norm_{i,w}}
\newcommand{\Rl}{R_{l}}
\newcommand{\Dqi}{D_{q,b}}
\newcommand{\Tizero}{T_{i,0}}

\newcommand{\Normi}{Norm_{i}}
\newcommand{\vbarycm}{v_{{\rm baryc},m}}
\newcommand{\vobsm}{v_{{\rm obs},m}}
\newcommand{\um}{u_{m}}
\newcommand{\Tequivstar}{T_{\rm{equiv,\star}}}
\newcommand{\SA}{S_{\rm{A}}}
\newcommand{\SN}{S_{\rm{N}}}

\newcommand{\Pn}{P_{n}}
\newcommand{\sigmav}{\sigma_{\rm{v}}}
\newcommand{\zu}{z_{\rm{u}}}

\newcommand{\sigmaewvm}{\sigma_{{\rm v,ew},m}}
\newcommand{\Rcount}{R_{{\rm count},m}}
\newcommand{\Raz}{R_{\mathrm{az},m}}
\newcommand{\Qlos}{Q_{\mathrm{LOS}}}
\newcommand{\QK}{Q_{\mathrm{K}}}
\newcommand{\QZ}{Q_{\mathrm{Z}}}
\newcommand{\QA}{Q_{\mathrm{A}}}
\newcommand{\Qw}{Q_{w}}
\newcommand{\Lm}{L_{m}}
\newcommand{\QL}{Q_{\mathrm{L}}}

\begin{document} 

\defcitealias{Bartalesi24}{B24}

   \title{Gas rotation and turbulence in the galaxy cluster Abell 2029}

   \author{T. Bartalesi\inst{1,2,3}\fnmsep\thanks{\email{tommaso.bartalesi3@unibo.it}}
          \and
          A. Simionescu\inst{3}
          \and
          S. Ettori\inst{2,4}
          \and
          C. Nipoti\inst{1}
          \and
          V. Ghirardini\inst{2}
          \and
          A. Sarkar\inst{5,6}
          \and 
          M. Sun\inst{7} 
          }


   \institute{Dipartimento di Fisica e Astronomia “Augusto Righi” – Alma Mater Studiorum – Università di Bologna, via Gobetti 93/2, I-40129 Bologna 
         \and INAF, Osservatorio di Astrofisica e Scienza dello Spazio, via Piero Gobetti 93/3, 40129 Bologna, Italy 
         \and SRON Netherlands Institute for Space Research, Niels Bohrweg 4, 2333 CA Leiden, the Netherlands
         \and INFN, Sezione di Bologna, viale Berti Pichat 6/2, 40127 Bologna, Italy    
         \and Department of Physics, University of Arkansas, 825 W Dickson st, Fayetteville, AR, USA 
         \and MIT Kavli Institute for Astrophysics and Space Research, 70 Vassar st., Cambridge, MA, USA
         \and Department of Physics and Astronomy, The University of Alabama in Huntsville, Huntsville, AL 35899, USA
             }

\date{Draft, December 19, 2025}

  \abstract
  {} 
   {
   We constrain the rotation and turbulent support of the intracluster medium (ICM) in Abell 2029 (A2029), using dynamical equilibrium models and a combination of state-of-the-art X-ray datasets.
   }
   {
   The rotating, turbulent ICM in the model has a composite polytropic distribution in equilibrium in a spherically-symmetric, cosmologically motivated dark halo. 
   The profile of rotation velocity and the distribution of turbulent velocity dispersion are described with flexible functional forms, consistent with the properties of synthetic clusters formed in cosmological simulations. 
   Adopting realistic profiles for the metallicity distribution of the ICM and for the point spread function of XRISM and XMM-Newton, we tune via a Markov chain Monte Carlo algorithm the observables of the intrinsic quantities of the plasma in our model to reproduce the radial profiles of the thermodynamic quantities as derived from the spectral analysis of the XMM-Newton and Planck maps and the measurements of the line-of-sight (LOS) non-thermal velocity dispersion and redshift (probing the LOS velocity) in the XRISM pointings. 
   }
    {
    Our model accurately reproduces the measurements of redshift and LOS non-thermal velocity dispersion, as further demonstrated by simulating and analyzing synthetic counterparts of the XRISM spectra, in accordance with the posterior distributions we obtain.
    We measure turbulence-to-total pressure ratio $\approx$ 2\% across the (0 -- 650) kpc radial range, and, under the hypothesis that rotation is the only bulk motion, a rotation-to-dispersion velocity ratio peaking at 0.15 between 200 -- 600 kpc.
    The hydrostatic-to-total mass ratio is $\approx$ 0.97 at $r_{2500}$, the radius enclosing an overdensity of 2500 times the average value. Further constraints on the presence and amount of rotation could be obtained through a full azimuthal coverage of A2029 with XRISM.}
    {}
   \keywords{Galaxies: clusters: general -- Galaxies: clusters: individual: Abell~2029 -- Galaxies: clusters: intracluster medium -- X-rays: galaxies: clusters                }

   \maketitle
%

\section{Introduction}
\label{sec:intro}
Improving the accuracy in the estimation of the mass of galaxy clusters would reduce the systematics in the determination of the parameters that define the cosmological background \citep[e.g.,][for a review]{Pratt19}.
Even though the matter content of galaxy clusters is dominated by dark matter, they are permeated by a hot, rarefied, optically thin plasma, known as the intracluster medium (ICM).
It manifests itself primarily in the X-ray band, via both the Bremsstrahlung and heavy-metal line emission \citep[e.g.,][for a review]{Sanders23}, and in the microwave band, via weak distortions in the black-body spectrum of the cosmic microwave background (CMB) caused by thermal and non-thermal motions of the ICM relative to the CMB rest frame \citep{Sunyaev72, Sunyaev80}.
A key step in improving the accuracy in the dynamical mass estimation from the observations of the ICM is to go beyond the hydrostatic-equilibrium assumption \citep[e.g.,][]{Pratt19}.
However, the modest energy resolution of X-ray CCD spectrometers, such as those abroad Chandra and XMM-Newton, and the limited sensitivity of microwave detectors have prevented us from characterizing the velocity field of the ICM, except for a small number of nearby clusters, for which the measurements are poorly precise \citep[e.g.,][]{sanders20, Gatuzz22, Gatuzz22b, Sanders23,Gatuzz24}.
The launch of the XRISM satellite\footnote{\url{https://www.xrism.jaxa.jp/en/}} \citep{Tashiro18}, following the short-lived Hitomi, marked the beginning of the era of high-resolution X-ray spectroscopy, capable of detecting shifts in the centroids and of the broadenings of the X-ray emission lines corresponding to ICM velocities down to $\sim 10\, \kms$ in the inner regions of galaxy clusters.
Thanks to the growing sample of clusters observed by Hitomi and XRISM \citep{Hitomi16, XRISM_centaurus25,XRISM_a2029_1_25, XRISM_coma25}, and the joint analysis of X-ray, SZ, and strong gravitational lensing data \citep{Sayers21, Chappuis25}, the low contribution of non-thermal motions to the ICM dynamics in their central regions is becoming increasingly solid evidence.
However, the poor signal-to-noise ratio in the available data has so far limited these types of studies to a small sample of nearby, massive clusters.

Some authors have attempted to address the specific question of whether and to what extent the ICM rotates.   \citet[hereafter \citetalias{Bartalesi25}]{Bartalesi25}, based on X-ray data of a sample of massive galaxy clusters, find that there is room for non-negligible rotation in the ICM, with a typical peak rotation speed $\approx 300 {\rm \,km\,s^{-1}}$. 
Velocities of the ICM also imprint weak distortions in the cosmic microwave background spectrum, known as kinetic Sunyaev-Zeldovich (kSZ) effect, which can be observed in the microwave band, in particular with the upcoming Simons observatory \citep{Abitbol25}. The rotational component of kSZ has been recently detected at $3.6\sigma$ confidence level through the analysis of the signal in the Planck images combined with the SDSS redshifts of the member galaxies in 25 nearby galaxy clusters \citep{Goldstein25}.
The kinematics of the ICM has long been studied in clusters formed in cosmological hydrodynamical simulations, where rotation and turbulence play a minor yet relevant role in supporting the ICM against the cluster gravity.
While the rotation of the ICM is expected to peak at intermediate radii \citep[e.g.,][]{BALDI17, Altamura23}, turbulence support is predicted to increase outwards \citep[e.g.][hereafter \citetalias{angelinelli20}]{lau13, nelson14, angelinelli20}.
The XRISM spectra have the unprecedented potential to test these predictions, although they probe only spatially limited regions of nearby clusters. Significant azimuthal variations are expected in real systems, similar to those formed in cosmological simulations \citep{nelson14}.

Abell~2029 (hereafter, A2029) is a massive, nearby cluster at a redshift $z_0 \approx 0.0779$ as inferred from the optical spectroscopy of the brightest cluster galaxy \citep[BCG;][]{XRISM_a2029_1_25}.
The peak of the ICM emission lies close to its centroid \citep{Eckert22b}, no relative displacement and motion between BCG and ICM is observed \citep{XRISM_a2029_1_25}, and the X-ray isophotes are well approximated by an elliptical shape that we estimate from the XMM-Newton soft images to have an emission-weighted average ellipticity of $\approx$ 0.08. Therefore, we can consider A2029 an exceptionally poorly disturbed cluster and, thus, a very close representation of a symmetric ICM in equilibrium within the potential well.
It was one of the targets of XMM Cluster Outskirts Project (X-COP), a large observational campaign aimed to characterize the profiles of the thermodynamic quantities of the ICM from the inner regions out to the outskirts by leveraging the joint analysis of the X-ray and Sunyaev-Zeldovich (SZ) data \citep[][]{X-COP}.
For A2029, this campaign yielded a high-quality dataset of density, temperature, and thermal pressure profiles extending from the center to the outskirts \citep[][hereafter \citetalias{Ghirardini19}]{Ghirardini19}, which has been exploited by \cite{Ettori19} and \cite{Eckert22} to derive hydrostatic-equilibrium models for A2029.
More recently, during the verification phase of the XRISM mission, A2029 was observed with the micro-calorimeter spectrometer XRISM/Resolve through three pointings along the North-West branch, with a total exposure of $\approx 500$ ks.
The spectra extracted from these three pointings revealed a low but still significant contribution from the non-thermal motions in the ICM, as compared to the temperatures measured from the same spectra \citep[][hereafter \citetalias{XRISM_a2029_2_25}]{XRISM_a2029_2_25}.
In this paper, we study the properties of A2029 as inferred by applying hydrodynamical-equilibrium models of the same family as those presented in  \citetalias{Bartalesi25}, which include rotation and turbulence of the plasma, to the data from the X-COP project and observed by XRISM/Resolve.

Given that XRISM is characterized by a large extent of the point spread function (PSF; $\approx$ 1.3 arcmin full-width-at-half-maximum), counts measured in the X-ray spectrum originate from photons emitted by the ICM in the plane-of-the-sky regions surrounding the pointing under analysis.
If not properly accounted for, this effect, known as spatial-spectral mixing (SSM), introduces potentially significant contamination to the physical information encoded in the spectra.
The most accurate, albeit computationally intensive, method to address this during the spectral forward-fitting involves adopting a thermal model with both spatial (in the plane of the sky) and spectral (in energy) dependencies, and simulating stochastically the paths of X-ray photons from the plane of the sky to the XRISM/Resolve focal plane. 
This technique, known as Monte Carlo spectral simulation, forward-folds the model through the instrument’s response characteristics.
In the case of $N$ adjacent pointings, it is standard to model the spectrum extracted from the $j$-th pointing as a combination of contributions from all $i$-th regions corresponding to the available pointings, where $i, j=\{1, ..., N \}$.
Assuming a Chandra broad-band surface brightness map (treated as energy-independent), we run Monte Carlo simulations to build a $N\times N$ matrix of specific ancillary response functions (ARFs), with the generic element $ARF_{i, j}$.
Multiplying the thermal emission model for the generic $i$-th region by $ARF_{i, j}$ yields an estimation of the fractional contribution of photons originating from the $i$-th region to the observed $j$-th spectrum.
Then, the $N$ spectra are simultaneously fit to infer the parameters of each $i$-th thermal model, primarily the shifts and broadening of the X-ray emission lines \citep[e.g.][and also \citetalias{XRISM_a2029_2_25}]{Hitomi18, XRISM_centaurus25}.
Several factors may complicate or limit this approach: for example, the poor signal-to-noise ratio in the spectra extracted from the outer pointings, the small number of X-ray emission lines detected with high significance \citep[in the outermost pointing of A2029 only three lines are detected at a 3$\sigma$ confidence level in the energy range 2 -- 10 keV;][]{Sarkar25} and the reduced contrast between emission lines and the continuum in the presence of a $\sim 100\, \kms$ line-of-sight (LOS) non-thermal velocity dispersion.
The fluxes of the X-ray emission lines, necessary for the ICM velocity measurements, are determined by a combination of emission measure, temperature, and metallicity of each fluid element contributing to the observed spectrum. 
Thus, an accurate model-data comparison should simultaneously account for all these effects.

Motivated by these issues, but limited by the huge computational cost of running a spectral simulation for every model evaluation in the fitting procedure, we adopt a two-step strategy.
(i) In the Bayesian framework of model-data comparison, we tune the average observable quantities of our model in the exposure regions to reproduce those inferred from the observational data from XMM, Planck and XRISM/Resolve.
(ii) From the optimized model, we generate via a Monte Carlo spectral simulation XRISM/Resolve-like observations of our A2029 model for the same pointings as for the real data. 
We, then, evaluate whether the synthetic XRISM/Resolve data reproduces the observed line shifts and broadenings as measured from the real data.
For the sake of consistency with the mock observations, we redo the spectral reduction and analysis of the XRISM/Resolve data.

The paper is organized as follows.
Sects.~\ref{sec:data} and \ref{sec:model} detail the observational data and the model used for our analysis, respectively. 
The method and results of the joint fitting to X-COP and XRISM/Resolve data are presented in Sect.~\ref{sec:joint}.
In Sect.~\ref{sec:simulation}, we generate and compare synthetic spectra with the real ones.
Sect.~\ref{sec:conclusion} summarizes our work and the main results obtained.
In this work, we assume a flat $\Lambda$ cold dark matter (CDM) cosmological model, with present-day matter density parameter $\Omega_{\mathrm{m},0}=0.3$ and Hubble constant $H_0=70\,\kms \mathrm{Mpc^{-1}}$.  
We define $M_\Delta$ as the mass enclosed within a sphere of radius $r_\Delta$ where the average density is $\Delta$ times the average density of the Universe, $\rhocrit$.
We use the {\tt Heasoft} package version 6.35, {\tt XSPEC} software version 12.0, with the default atomic database {\tt AtomDB}, and the calibration files of XRISM/Resolve from {\tt CalDB v8} repository.


\section{Observational data}
\label{sec:data}
This section outlines the observational datasets that we will compare to our model in our analysis of A2029.
Sect. \ref{sec:resolve} describes the reduction and analysis of the X-ray high-resolution spectra obtained with XRISM/Resolve. 
Sect. \ref{sec:XMM} details the measurements reported by \citetalias{Ghirardini19} as part of the X-COP project and obtained from XMM-Newton observations at modest spectral resolution in the X-rays and from modeling of the SZ effect signal from Planck data.

\subsection{Resolve data}
\label{sec:resolve}
During the verification phase of the XRISM mission, A2029 was observed by XRISM/Resolve with three $(3\, \mathrm{arcmin})\times (3\, \mathrm{arcmin})$ pointings (approximately, 270 kpc$\times$270 kpc at $z_0 = 0.0779$) aligned along a contiguous arm in the North-East (NE) direction out to 668 kpc, which is close to $r_{2500} \approx 680$ kpc \citepalias[see Figure 1 of][]{XRISM_a2029_2_25}. We refer to these pointings with an index $m = \{1, 2, 3\}$, where $m = 1$ indicates the central one.

We begin the data reduction process using the cleaned event files. 
We exclude events from pixel 12 (used for gain reconstruction) and pixel 27 \citepalias[which exhibited unexpected scale jumps; see][for details]{XRISM_a2029_2_25}. 
We apply the standard screening procedure recommended in {\tt xapipeline} and implemented in the {\tt Heasoft} software.
We extract the high-resolution primary photon counts integrated over all remaining detector pixels, producing the spectral file for the $m$-th pointing.
Using archival night Earth data and applying the same screening, we generate the non X-ray background NXB$_m$ using {\tt rslnxbgen}. 
Next, we build a new screened event file by excluding all the events of the $m$-th spectral file with PI values outside the 6000–20000 range.
From this file, we generate the Redistribution Matrix File (RMF$_m$) using {\tt rslmkrmf}\footnote{The procedure we follow to generate RMF$_m$ is detailed at the link \url{https://heasarc.gsfc.nasa.gov/docs/xrism/analysis/ttwof/index.html}.} and the ARF$_m$ of a point-like source being in the center of the XRISM/Resolve Field of View (FoV) using {\tt xaarfgen}.
Given that the X-ray background is significantly lower than the thermal emission in all the observations, as evidenced in Fig. 2 of \citetalias{XRISM_a2029_2_25}, we neglect it in our analysis.
As discussed in Sect. 2 of \citetalias{XRISM_a2029_2_25}, for the $m$-th pointing the systematic uncertainties on the position of the emitting lines in the (6 -- 7) keV range and their broadening is measured to be equal to or lower than $\pm$ 0.15 eV and $\pm 6$ $\kms$, respectively. 

We begin the spectral analysis for the generic $m$-th pointing with its corresponding reduced spectrum.
Using the {\tt grppha} task of {\tt Heasoft}, we bin it such to ensure at least 1 count per bin in each bin\footnote{Given that \citetalias{XRISM_a2029_2_25} binned the reduced spectrum using the optimal binning algorithm of \cite{Kaastra17} as implemented in the {\tt ftgrouppha} task of the {\tt Heasoft} package, we verified that the difference in the inferred values of $\zum$ and $\sigmavm$ (see below) between these two binnings does not exceed the corresponding 1$\sigma$ statistical error.}. 
In {\tt XSPEC} we subtract from each of them  the corresponding non X-ray background with the command {\tt backgrnd}.
We account for the galactic absorption with the model {\tt tbabs} \citep{Wilms00}, fixing the column density, $\Nh$, to $3\times 10^{20} \mathrm{cm^{-2}}$ as recommended by \citetalias{XRISM_a2029_2_25}. 
We, then, fit the $m$-th spectrum in the energy range $3 - 10$ keV with the thermal emission model for a uniform and homogeneous plasma in collisional-ionization equilibrium (CIE), 
{\tt bapec}\footnote{\url{https://heasarc.gsfc.nasa.gov/xanadu/xspec/manual/XSmodelApec.html}} \citep{Smith01}. 
Given that no significant parameter correlations are expected as shown by the spectral simulations of \cite{Bartalesi24} \citepalias[hereafter,][see their Sect. 5.2]{Bartalesi24}, we estimate their $1\sigma$ errors obtained from the diagonal terms of the covariance matrix, using the Levenberg-Marquardt algorithm as implemented in {\tt XSPEC}.
Table \ref{tab:resolve} reports the measurements of the redshift, $\zum$, and LOS non-thermal velocity dispersion, $\sigmavm$, for the three pointings ($m = 1, 2, 3$), which we compare to the model in Sect. \ref{sec:joint}.
The velocity of the uniform and homoegeous plasma with respect to $z_0$ in the $m$-th pointing, $\um$ (positive for receding ICM), and the barycentric velocity of the Earth, $\vbarycm$, determine $\zum$, according to  \citep[e.g.,][]{Roncarelli18}
\begin{equation}
     \zum = (1 + z_0) \sqrt{(1 - \vobsm/c) / (1 + \vobsm / c)} - 1.
     \label{eq:z_u}
\end{equation}
where $c$ is the light speed and $\vobsm = \vbarycm + \um$, with $\vbarycm = 26$ $\kms$ for $m = \{2, 3\}$ and $v_\mathrm{baryc,3} = -27$ $\kms$.
As discussed in Sect. 2 of \citetalias{XRISM_a2029_2_25} (see their Table 3), the plasma in the $m=\{2, 3 \}$ spectra is significantly blueshifted with respect to $z_0$ (see also the differences between $\zum$ and $z_0$ in Table \ref{tab:resolve} for $m=\{2, 3 \}$). 
When we compare our model with the Resolve data, we therefore assume approaching (i.e., negative) rotation velocities of the plasma.

\begin{table}
      \caption[]{Measurements from the XRISM/Resolve spectra.}
     $$ 
         \begin{array}{llll}
            \hline
            \noalign{\smallskip}
            \mathrm{Pointing/OID} & m & \zum[10^\mathrm{-2}] & \sigmavm [\kms] \\
            \noalign{\smallskip}
            \hline
            \noalign{\smallskip}
            \mathrm{Center}/000151000 & 1 & 7.775 \pm 0.005 & 157 \pm 11\\
            \mathrm{N1}/000150000 & 2 & 7.728 \pm 0.010 & 135 \pm 19\\
            \mathrm{N2}/300053010 & 3 & 7.755 \pm 0.013 & 140 \pm 33\\
            \hline
         \end{array}
     $$ 
     \tablefoot{From left to right: name and index of the pointing, measured redshift and LOS non-thermal velocity dispersion. 
     The statistical errors are reported to $1\sigma$ confidence.}
     \label{tab:resolve}
\end{table}

\subsection{Data from XMM and Planck observations}
\label{sec:XMM}
We now present the SZ data and X-ray spectral results for A2029 obtained from the spectral analysis of \citetalias{Ghirardini19} and made available on the X-COP website\footnote{\url{https://dominiqueeckert.wixsite.com/xcop/about-x-cop}
}.

A2029 was observed with XMM-Newton through multiple pointings forming a mosaic. 
Assuming the cluster center to coincide with the peak of the X-ray surface brightness map, the mosaic covers a symmetric region around the cluster center, extending out to a projected radial distance of approximately 1488 kpc.
\citetalias{Ghirardini19} extracted 14 X-ray spectra, each from a circular annulus indexed with $i$, with inner and outer radii $\Ri$ and $\Rplus$, respectively \citepalias[see Sect. 2.4][for details]{Ghirardini19}.
Each spectrum was fitted using the {\tt apec} emission model (a version of {\tt bapec} without LOS non-thermal velocity broadening; see Sect. \ref{sec:resolve}; \citealt{Smith01}), yielding measurements of the normalization of the thermal continuum, $\Normi$, the temperature, $\Ti$, and the metallicity, $\Zi$, of an uniform and homogeneous plasma in CIE. 
\citetalias{Ghirardini19} also computed the brightness
\begin{equation}
    \label{eq.NNN}
    \Niii = \frac{\Normi}{\pi \left[ \left(\Rplus / [\arcminute] \right)^2 - \left(\Ri / [\arcminute] \right)^2 \right]}.
\end{equation}
The measurements of $\Niii$ and $\Ti$ taken from the X-COP website are compared to our model in Sect. \ref{sec:joint}\footnote{The spectral analyses of XMM-Newton and XRISM/Resolve data provide independent constraints on the normalization of the thermal continuum (proportional to the density squared) and the temperature of the ICM in the inner regions of A2029. 
However, given that both quantities may be subject to uncertainties in the cross-calibration of the telescope effective areas, the instrumental PSF and in the NXB modeling, in the comparison with our model, we adopt only the XMM-Newton measurements, which additionally offer more complete azimuthal coverage of the cluster in the plane of the sky (see text). 
Sect. \ref{sec:results} discusses in part the consistency of the properties of the plasma in our model, which reproduces the XMM spectral results, with the XRISM/Resolve spectral data.}.

The Planck observation of A2029 results in a SZ signal map that resembles a square box in the plane of the sky, centered at the center of A2029 and with a side length of 7 Mpc. 
\citetalias{Ghirardini19} modeled the SZ signal map, under the assumption that the thermal SZ effect dominates the signal (i.e., the relation between the "SZ pressure" and the Compton-parameter in Eq. 3 of \citetalias{Ghirardini19} holds).
Assuming spherical symmetry of the ICM distribution and adopting a realistic profile for the azimuthally-averaged Planck PSF, they performed a geometric deprojection of the SZ signal, obtaining measurements of the "SZ pressure" in ten concentric spherical shells (see Sect. 2.5 of \citetalias{Ghirardini19}, for details).
Given that, as discussed in detail in \citetalias{Ghirardini19}, the innermost three shells have radial widths smaller than the full-width-at-half-maximum of the Planck PSF, we exclude them from our analysis.
In the X-COP website, a covariant matrix accounts for the statistical errors and the cross-correlations between the values of $\PSZj$ at all $j$.
Given that the diagonal terms of the covariance matrix dominate, we include only these diagonal terms in the error budget of these measurements.
In Sect. \ref{sec:joint}, we compare our model to the measurements of the thermal "SZ pressure", $\PSZj$, in seven spherical shells, indexed by $j=\{1,…,7\}$ and ranging from $\approx 800$ kpc out to $3.5$ Mpc.

\section{Model of pressure-supported, rotating ICM}
\label{sec:model}
Sections~\ref{sec:intrinsic} and \ref{sec:from_to} detail, respectively, the intrinsic and observable properties of the axisymmetric cluster model with turbulent and rotating plasma used in our analysis.
The turbulence is assumed to be isotropic and the rotation is such that the surfaces of constant angular velocity are cylinders.

\subsection{Intrinsic properties of the model}
\label{sec:intrinsic}

We adopt a cylindrical reference frame ($R$, $\phi$, $z$), with the origin at the minimum of the gravitational potential $\Phi$.
As in \citetalias{Bartalesi25}, the self-gravity of the plasma is neglected.
$\Phi(r)$, with $r=\sqrt{R^2 + z^2}$ the spherical radius, is a spherically-symmetric Navarro-Frenk-White \citep[NFW; ][]{Navarro96} potential, fully specified by the virial mass $\Mtwo$ and concentration $c_{200}$.
The rotation curve of the plasma is \citep[e.g.][]{Bianconi13, Bartalesi25}
\begin{equation}
    \label{eq:rotation}
    u_\phi(S) = 4\upeak \frac{S}{(1 + S)^2},
\end{equation}
where $S = R / \Rpeak$, $\upeak$ and $\Rpeak$ are the peak rotation speed and its radius, respectively.
\cite{BALDI17} showed that this functional form reproduces the average rotation speed profile of the ICM in clusters formed in a cosmological simulation.
The full expression for the effective potential $\Phieff(R, z)$ in our model is provided in equations (24) and (25) of \citetalias{Bartalesi24}.

The plasma number density and the total (thermal plus turbulent) pressure of a polytropic distribution in equilibrium in the axisymmetric potential $\Phieff$ are \citep[e.g.,][]{Tassoul78}
\begin{equation}
    \label{eq:n_0}
    n(R, z) = n_0 \left\{1 - \frac{\gamma' - 1}{\gamma'} \frac{\rho_0}{P_0} \left[\Phieff(R, z) - \Phieffzero \right] \right\}^{\frac{1}{\gamma' - 1}}
\end{equation}
and
\begin{equation}
    \label{eq:p_0}
    P(R, z) = P_0 \left( \frac{n(R, z)}{n_0} \right)^{\gamma'},
\end{equation}
respectively, where $\Phieffzero = \Phieff(R_0, 0)$, $n_0 = n(R_0, 0)$, $P_0 = P(R_0, 0)$, $\gamma'$ is the polytropic exponent and $R_0$ is a generic radius in the equatorial plane. 
Here, $\rho_0 = \rho(R_0, 0)$, where $\rho = n \mu \proton $ is the gas mass density.
$\proton$ is the proton mass and $\mu = 0.6$ the mean molecular weight, assuming a metallicity of 0.3 times the solar value throughout the cluster.

We consider a composite polytropic distribution with three distinct polytropic exponents: $\gamCEN$, $\gamIN$ and $\gamOUT$, such that $\gamCEN \leq \gamIN \leq \gamOUT$.
The transitions occur at radii $\Rcen$ (from $\gamCEN$ to $\gamIN$) and $\Rstar$ (from $\gamIN$ to $\gamOUT$) in the equatorial plane.
In Eqs. \ref{eq:n_0} and \ref{eq:p_0}, the reference potentials for the transitions in the meridional plane are $\Phieffzero = \{\PhiCEN, \Phistar \}$ for $R_0 = \{\Rcen, \Rstar\}$, respectively, where $\PhiCEN = \Phieff(\Rcen, 0)$ e $\Phistar = \Phieff(\Rstar, 0)$. 
The corresponding reference plasma properties are $n_0 = \nstar$ and $P_0 = P_\star$ for $R_0 = \Rstar$, and $n_0 = n(\Rcen, 0)$ and $P_0 = P(\Rcen, 0)$ for $R_0 = \Rcen$.
Accordingly, $\gamma'$ takes the value of $\gamCEN$, $\gamIN$ or $\gamOUT$ for $\Phieff(R, z) < \PhiCEN$, $\PhiCEN \leq \Phieff(R, z) < \Phistar$ and $\Phieff(R, z) \geq \Phistar$, respectively.
The two-exponent composite-polytropic distribution of \citetalias{Bartalesi24} and \citetalias{Bartalesi25} is recovered in the limit case $\gamCEN = \gamIN$.

As in \citetalias{Bartalesi25}, the turbulent pressure is $\pturb = P - n \kboltz \Tgas = \alphaturb P$, where $\Tgas$ is the plasma temperature and $\kboltz$ the Boltzmann constant. 
The distribution of the turbulent-pressure-to-total-pressure ratio follows
\begin{equation}
    \label{eq:alpha_turb}
    \alphaturb(P) = (\alphainf - \alphazero) \frac{\ln \left[1 + P / (\xi P_\star) \right]}{P / (\xi P_\star)} + \alphazero.
\end{equation}
where $\alphazero$ and $\alphainf$ are, respectively, the asymptotic values for $P \to \infty$ and $P \to 0$, and $\xi$ defines the transition between $\alphazero$ and $\alphainf$.
This expression can provide a good representation of the median profile of the turbulent support of the ICM in clusters formed in the cosmological simulation analyzed by \cite{angelinelli20} \citepalias[see Sect. 3.1 of][]{Bartalesi25}.

We define the parameter $\Tstar \equiv P_\star / (\nstar \kboltz)$ as the equivalent temperature associated with the total pressure support at the reference point ($\Rstar$, 0) in the meridional plane.
This model has 14 parameters: $\gamIN$, $\gamOUT$, $\gamCEN$, $\Rstar$, $\Rcen$, $\nstar$, $\Tstar$, $\Mtwo$, $c_{200}$, $\upeak$, $\Rpeak$, $\xi$, $\alphainf$ and $\alphazero$.

\subsection{From the intrinsic to the observable quantities of the model}
\label{sec:from_to}
We adopt a Cartesian reference system ($x$, $y$, $z$), with the origin located at the minimum of $\Phi$. 
In our model, this point coincides with the point with the highest value of the plasma density.
We fix the redshift of the point with the highest plasma density in our model to $z_0$.
We assume that our axisymmetric model is observed with the symmetry and rotation axis $z$ in the plane of the sky, i.e.\ in the configuration that maximizes the component of the rotation speed of the plasma along the LOS. 
The $x$-axis is chosen to lie along the LOS.
The $y$-axis in the plane of the sky is oriented along the NE--SW direction of A2029 \citepalias[see Fig. 1 of][]{XRISM_a2029_2_25}, i.e.\ along the direction sampled with the Resolve pointings, with positive values in the NE sector.
For the sake of simplicity, we approximate the FoV of Resolve as a square with the two perpendicular sides of length 3 arcmin, parallel to either the $y$-axis or the $z$-axis, and with centers in the plane-of-the-sky points (0, 0), (3 arcmin, 0) and (6 arcmin, 0) for $m=\{1, 2, 3\}$, respectively.
In this simplified configuration, only the $m = 0$ pointing is symmetric with respect to the rotation axis $z$.

Let us consider a generic quantity $Q(R, z)$ dependent only on the intrinsic quantities in the model, as described in Sect. \ref{sec:intrinsic}.
We now introduce the polar coordinates in the $y$-$z$ plane $\Rhat = \sqrt{y^2 +z^2}$ and $\phihat = \arctan(z / y)$.
In our analysis, we use only quantities evaluated at $z=0$.
The integral of $Q(R,0)$ along the LOS is
\begin{equation}
    \label{eq.ffff}
    \Qlos(\Rhat) = 2 \int_{0}^{\infty} \diff x Q(R, 0),
\end{equation}
where $R = \sqrt{x^2 + y^2}$.
Hereafter, unless stated otherwise, we approximate the projected quantities of the model as circularly symmetric, with profiles $\Qlos(\Rhat)$.

In Appendix \ref{sec:PSF}, we describe how we convolve $\Qlos(\Rhat)$ with a circularly symmetric PSF profile, to obtain $\QA(\Rhat)$, when we approximate the evaluation of the PSF profile at $z=0$, and $\QK(\Rhat)$ in the general case.
When $\QA(\Rhat)$ is weighted over a realistic circularly symmetric projected metallicity profile of A2029, as detailed in Appendix \ref{sec:metals}, we obtain $\QZ(\Rhat)$.

Let us consider the $i$-th circular annulus in the plane of the sky, with inner and outer radii $\Ri$ and $\Rplus$, respectively.
The integral of the quantity $\Qw(\Rhat)$, with $w = \{0, K, A\}$, in the $i$-th circular annulus is \citepalias[see Eq. (D.3) of][]{Bartalesi25}
\begin{equation}
    \label{eq.Ii}
    \Iiw[Q] = 2\pi \int_{\Ri}^{\Rplus} \Rhat \diff \Rhat \Qw(\Rhat),
\end{equation}
where $\Qlos$, $\QK$ and $\QA$ are defined above.

Now, we consider in the plane of the sky the $m$-th rectangle with a side of length $\yup - \ylow$ parallel to $y$-axis and the other of length $\zup = 1.5$ arcmin (fixed for all $m$) parallel to the $z$-axis.
For $m = \{ 1, 2, 3\}$, we define $\ylow = \{0, 1.5, 4.5\}$ arcmin and $\yup = \{1.5, 4.5, 7.5\}$ arcmin, respectively (at $z_0$, 1 arcmin $\approx 88$ kpc). 
The integral of $\QZ$ (see Sect. \ref{sec:metals}) in the $m$-th rectangle is
\begin{eqnarray}
    \label{eq:SQm}
    \Sm[Q] = \int_{0}^{\zup} \diff z \int_{\ylow}^{\yup} \diff y \QZ \left(\sqrt{y^2 + z^2}\right).
\end{eqnarray}

The analogue of the normalization of the thermal continuum measured from our model in the $i$-th circular annulus in the plane of the sky, with inner and outer radii $\Ri$ and $\Rplus$, respectively, is
\begin{equation}
    \label{eq.Norm}
    \Normiw = C \Iiw\left[\nH \nee\right],
\end{equation}
where $\nH$ and $\nee$ are the hydrogen-equivalent and electron number densities of the plasma.
Given that metallicity gradients have a negligible impact on the plasma distribution in equilibrium in the gravitational potential, we simplify the model by assuming a constant metallicity of 0.3 times the solar value.
This implies $\nH = 1.17 \nee$ and $\nee = n / 1.94$, with $n$ obtained from Eq. (\ref{eq:n_0}).
In our model, we assume $n(R, z) = 0$ for $x, y, z >\RSZ$, where $\RSZ$ is the outer radius of the outermost spherical shell in which the SZ data have been extracted.
$C=10^{-14} / \left [ 4\pi \Dang^2 (1+z_0)^2 \right]$, with $\Dang$ the angular distance of the cluster, is the normalization factor adopted in {\tt XSPEC}.
As accurate approximation to the spectroscopically measured temperature in the presence of a non-neglibible temperature gradient in the ICM, \cite{Mazzotta04} proposed the spectroscopic-like temperature, which is measured from our model in the $i$-th circular annulus in the plane of the sky as
\begin{equation}
    \label{eq.Tsl}
    \Tiw=\frac{\Iiw\left[\nH \nee \Tgas^{3/4}\right]}{\Iiw\left[\nH \nee \Tgas^{-1/4}\right]},
\end{equation}
where $\Tgas$ is obtained as described in Sect. \ref{sec:intrinsic}.
In a set of high-resolution X-ray spectral simulations for a cluster formed in a cosmological simulation, \cite{Roncarelli18} found the emission-weighted LOS velocity and velocity dispersion as accurate approximations of the corresponding measurements inferred with the {\tt bapec} model.

We now consider the velocity vector of the plasma in our axisymmetric model, defined at a generic point $(x, y, z)$ as $\vec u = u_x (R, \phi) \hat e_x + u_y(R, \phi) \hat e_y$, where $u_x = u_\phi (R) \cos \phi$, $u_y = u_\phi(R) \sin \phi$ and $u_\phi$ is the rotation speed of the plasma in Eq. (\ref{eq:rotation}). 
$\hat e_x$ and $\hat e_y$ are the unit vectors defining the $x$ and $y$ directions.
In our configuration for the model (see above), the LOS component of $\vec u$ corresponds to $u_x$, which for $y \geq 0$ is equal to $u_\phi  \Rl/ R$, where $\Rl = \{\Rhat, \Rbar\}$ for $w = \{0, A\}$, respectively.
The emission-weighted LOS rotation speed measured from our model in the $i$-th circular annulus is 
\begin{equation}
    \label{eq.uew}
     \uiw = \frac{\Iiw[\nH \nee u_\phi  \Rl/ R]}{\Iiw[\nH \nee]}.
\end{equation}
When measured from our model in the $m$-th rectangle in the plane of the sky, the emission-weighted LOS rotation speed is
\begin{equation}
    \label{eq.ulosm}
     \uewm = \frac{\Sm[\nH \nee u_\phi \Rbar / R]}{\Sm[\nH \nee]}.
\end{equation}

In addition to the thermal velocity dispersion accounted for in the {\tt bapec} emission model, the contributions at any $(x, y, z)$ in our axisymmetric model to the LOS velocity dispersion evaluated at $(y, z)$ come from the one-dimensional (1D) turbulent velocity dispersion, $\sigmaturb(R, z) = \sqrt{\alphaturb P / (n \mu \proton)}$ and the spread in $u_x(R, \phi)$ along the $x$-axis. 
The emission-weighted non-thermal LOS velocity dispersions of the plasma as measured from our model in the $i$-th circular annulus and in the $m$-th rectangle in the plane of the sky are
\begin{equation}
    \label{eq.sigmaew}
     \sigmavi = \frac{\Iizero \left[\nH \nee \sqrt{\sigmaturb^2 + \left(u_\phi \Rhat / R - u_{i,0} \right)^2} \right]}{\Iizero \left[\nH \nee \right]}
\end{equation}
and
\begin{equation}
    \label{eq.sigmam}
    \sigmaewvm = \frac{\Sm \left[\nH \nee \sqrt{\sigmaturb^2 + \left(u_\phi \Rbar / R - u_{i,0}\right)^2} \right]}{\Sm[\nH \nee]},
\end{equation}
respectively.

The $\PSZj$ in the $j$-th spherical shell, with inner and outer radii $\radj$ and $\radplus$, respectively, is evaluated in our model following Sect. 3.2 of \citetalias{Bartalesi25}:
\begin{equation}
    \label{eq.model_press}
    \PSZj = \eta \frac{3 \int_{\radj}^{\radplus} y^2 \diff y\, \nee(|y|, 0) \kboltz \Tgas(|y|, 0) }{\radplus^3 - \radj^3},
\end{equation}
where $\eta$ is a dimensionless parameter accounting for the possible systematic offset between the electron pressure of the ICM and that measured from the Compton $y$-parameter maps.

\section{Joint fitting of X-ray data from XMM and Resolve and of SZ data from Planck}
\label{sec:joint}
Our model, presented in Sect. \ref{sec:model}, is tuned to reproduce via a Markov Chain Monte Carlo (MCMC) algorithm the measurements of the normalization of the thermal continuum, spectroscopic temperature, SZ pressure, redshift and LOS non-thermal velocity dispersion obtained from XMM, Planck and XRISM data (see Sect. \ref{sec:data}).
Sect. \ref{sec:stat} describes the statistical method, while Sect. \ref{sec:posterior} presents the results.

\subsection{Statistical method}
\label{sec:stat}

Assuming $K(d)$ for XMM in Eq. (\ref{eq.conv}) (see Sect. \ref{sec:PSF}) and using Eqs. (\ref{eq.Norm}), (\ref{eq.Tsl}) and (\ref{eq.model_press}), we evaluate $Norm_{i,\mathrm{K}}$, $\Ti = T_{i,\mathrm{K}}$ and $\PSZj$ of the plasma in the model in the $i$-th circular annulus in the plane of the sky and $j$-th spherical shell, respectively (we recall that the normalization of the $\PSZj$ profile is regulated by $\eta$, defined in Sect. \ref{sec:from_to}).
Using Eq. (\ref{eq.NNN}) with $\Normi = Norm_\mathrm{i, K}$, we then compute $\Niii$ from our model in the $i$-th circular annulus.
Given the axial symmetry of the model, we limit the evaluation of $\sigmavm$ and $\zum$ to $y\geq 0$.
Assuming $K(d)$ for XRISM in Eq. (\ref{eq.f_metal}) and using Eqs. (\ref{eq.ulosm}) and (\ref{eq.sigmam}), we evaluate $\uewm$ and $\sigmaewvm$ from the model in the $m$-th rectangular region, respectively.
Given that the $m = 1$ integration region in the $y$-$z$ plane is symmetric with respect to the $z$-axis (which in our model is assumed to be the rotation axis), the rotation manifests itself as a contribution to the LOS velocity dispersion.
This implies $z_\mathrm{u,1} = z_0$ and $\sigma_\mathrm{v,1} = \sqrt{ \sigma_\mathrm{v,ew,1}^2 + u_\mathrm{ew,1}^2}$.
In the $m = \{2, 3\}$ pointings, which are not symmetric with respect to the $z$-axis, the rotation of the plasma in our model manifests itself as a shifting of the X-ray emission lines measured via the redshift $\zum$ that we obtain from Eq. (\ref{eq:z_u}) with $\um = -\uewm$. 
In these cases, $\sigmavm = \sigmaewvm$. 

The statistical errors, $\sigmastat$, on the observed $\Niii$, $\Ti$ and $\PSZj$ are obtained as the average between the upper and lower $1 \sigma$ errors reported in the X-COP website. 
The statistical errors on the observed $\zum$ and $\sigmavm$ are taken from Table \ref{tab:resolve}.
Following \citetalias{Bartalesi25}, we introduce three parameters, $\epsN$, $\epsT$, and $\epsSZ$, to account for a systematic contribution, $\epsilon \sigmastat$ with $\epsilon = \{ \epsN, \epsT, \epsSZ \}$, to the uncertainty in the measurements of $\Niii$, $\Ti$ and $\PSZj$\footnote{These nuisance parameters cover a wide range of possible systematics, related both to intrinsic source properties and instrument calibration. These include, for instance, the presence of unresolved clumps in the XMM maps as well as uncertainties in the PSFs of XMM and Planck.}, respectively.  
$\epsN$, $\epsT$ and $\epsSZ$ are assumed to have the same values in all the radial bins.
We denote each datum, whether $\Niii$, $\Ti$, $\PSZj$, $\zum$ and $\sigmavm$, with $\Dqi$ where $q = \{ 1, 2, 3, 4, 5 \}$, respectively, and $b$ indicates the spatial bin corresponding to this datum.
We assumed the generic $\Dqi$ to have a normal distribution with median equal to the corresponding value computed from the model and standard deviation $\sqrt{1 + \epsilon^2}  \sigmastat$ for $q = \{1, 2, 3 \}$ and $\sigmastat$ for $q = \{ 4, 5 \}$.
It follows that the generic likelihood of the datum $\Dqi$, $\mathcal L (\Dqi)$, is a normal distribution with the same median and standard deviation as described above.

\begin{table}
      \caption[]{Lower and upper bounds of the uniform prior distributions of the MCMC parameters and their marginal posterior credible interval.}
     $$ 
         \begin{array}{lll}
            \hline
            \noalign{\smallskip}
            \mathrm{Parameter} & \mathrm{prior \, range} & \mathrm{posterior \, inference}\\
            \noalign{\smallskip}
            \hline
            \noalign{\smallskip}
            \gamIN & [0.8, 1.0] & 0.96 \pm 0.02 \\
            \gamOUT & [1.1, 1.4] & 1.23 \pm 0.04\\
            \nu_\star & [-1.0, 2.0] & 0.2 \pm 0.1\\
            \Tequivstar /\mathrm{keV} & [6.0, 11.0] & 8.37 \pm 0.19\\
            \Rstar /\mathrm{kpc} & [300, 750] & 500 \pm 50\\
            \Mtwo / (10^{15} M_\odot) & [0.5, 1,4] & 1.06 \pm 0.05\\
            \ln c_{200} & [0.8, 2.5] & 1.68 \pm 0.04 \\
            \upeak / (\kms) & [50, 350] & 195 \pm 23 \\
            \Rpeak / \mathrm{kpc} & [150, 650] & 391 \pm 145\\
            \log \xi & [0.5, 2.5] & --\\
            \alphainf & [0, 0.2] & 0.01 \pm 0.01\\
            \alphazero & [0, 0.2] & 0.01\\
            \Rcen & [1, 150] & 98 \pm 11\\
            \gamCEN & [0.5, 0.9] & 0.71 \pm 0.02\\
            \eta & [0.5, 1.5] & 0.96 \pm 0.1\\
            \log \epsN & [-1.5, 1.2] & -(0.4 \pm 0.2)\\
            \log \epsT & [-1.5, 1.2] & -(0.9 \pm 0.3)\\
            \log \epsSZ & [-1.5, 1.2] & -(1.2 \pm 0.3)\\
            \noalign{\smallskip}
            \hline
         \end{array}
     $$ 
     \tablefoot{First column: parameter. Second column: range between lower and upper bounds of the uniform prior distribution. Third column: median value, with the corresponding error. As standard, we leave empty the latter column, if the posterior inference is not significant with respect to the prior.}
     \label{tab:prior}
\end{table}

The 18 free parameters in the MCMC sampling are $\gamIN$, $\gamOUT$, $\nu_\star \equiv \log (\nstar / [10^{-3} \mathrm{cm^{-3}}])$, $\Tequivstar$, $\Rstar$, $\Mtwo$, $\ln c_{200}$, $\upeak$, $\Rpeak$, $\log \xi$, $\alphainf$, $\alphazero$, $\eta$, $\log \epsN$, $\log \epsT$ and $\log \epsSZ$ (see Sect.~\ref{sec:intrinsic}, for the definition of the model parameters).

We assume uniform prior distributions on all these parameters within the ranges listed in Table \ref{tab:prior}.
We briefly comment on the lower bound of the prior on the radius $\Rpeak$ where the rotation curve peaks.
There is a growing consensus that turbulence is the main driver of the re-acceleration of relativistic electrons, which manifests observationally in A2029 as a radio mini-halo, an extended synchrotron emission detected at low radio frequencies and in the inner regions \citep{Govoni09, Murgia09}. 
Based on this interpretation, we expect a significant turbulent contribution to the observed $\sigma_\mathrm{v,1}$, although also the rotation can contribute to $\sigma_\mathrm{v,1}$ (see above).
We realize this scenario in our model by conservatively setting the lower bound of the uniform prior on $\Rpeak$ approximately to the radial extent of the first Resolve pointing ($\approx$ 150 kpc).

The MCMC algorithm maximizes the logarithmic posterior distribution of the generic parameter vector $\vec \theta = (\theta_1, ..., \theta_\mathrm{18})$: $\ln \mathcal P (\vec \theta) = \sum_{q = 1}^{5} \sum_{b} \ln \left[\mathcal L \left(\Dqi \right)\right]$, if all the values $\theta_1, ..., \theta_\mathrm{18}$ are within the corresponding prior ranges, and $\ln \mathcal P = -\infty$, otherwise.

\subsection{Results from the posterior distribution of our model}
\label{sec:posterior}
When the MCMC algorithm converges to a stationary posterior\footnote{Using the same sampler as in \citetalias{Bartalesi25}
(see their Appendix B, for details), we run the MCMC algorithm for 10000 iterations with 500 chains. Empirically, we find that the posterior becomes stationary after 3000 iterations, but we conservatively assume that the iteration 5000 is the endpoint of the burn-in phase. The posterior sampling is obtained by selecting all the $\vec \theta$ for all the chains every 700 iterations, starting from the endpoint of the burn-in phase.}, we draw the posterior sampling that we use for our analysis. 
The relative frequency of the values of the parameters in the posterior sampling can be visualized in the corner plot presented in Appendix \ref{sec:corner}.
The credible interval obtained from the marginal posterior sampling of each parameter is reported in Table \ref{tab:prior}.
For the analysis of the results of the MCMC run, from the posterior sampling we randomly extract 100 parameter vectors that we consider as a representative set of the posterior distribution of our model.

\begin{figure*}
   \centering
   \hbox{
   \includegraphics[width=0.49\textwidth]{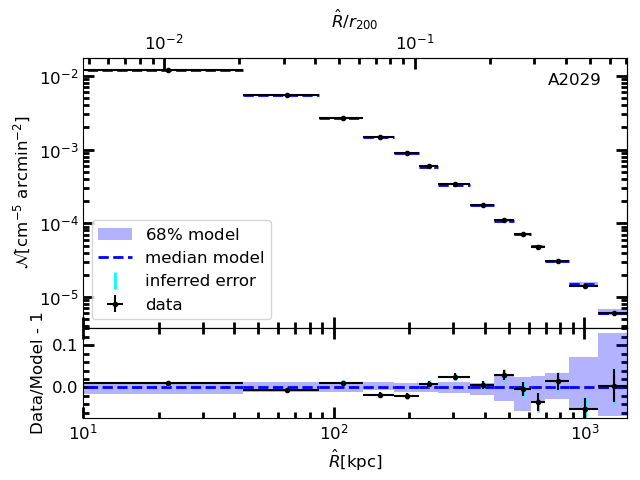}
   \includegraphics[width=0.49\textwidth]{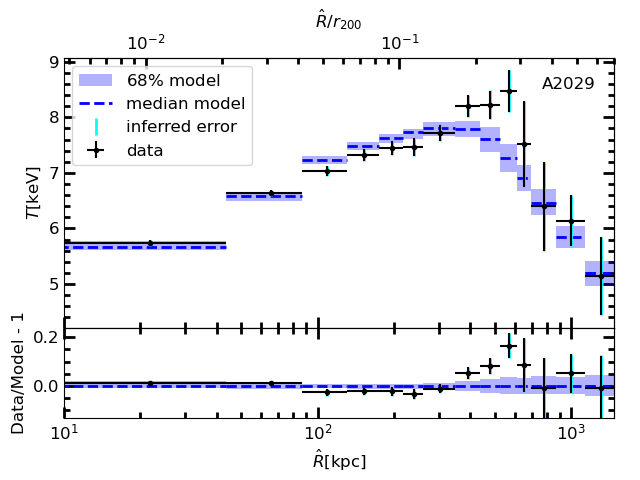}}
   \hbox{
   \includegraphics[width=0.49\textwidth]{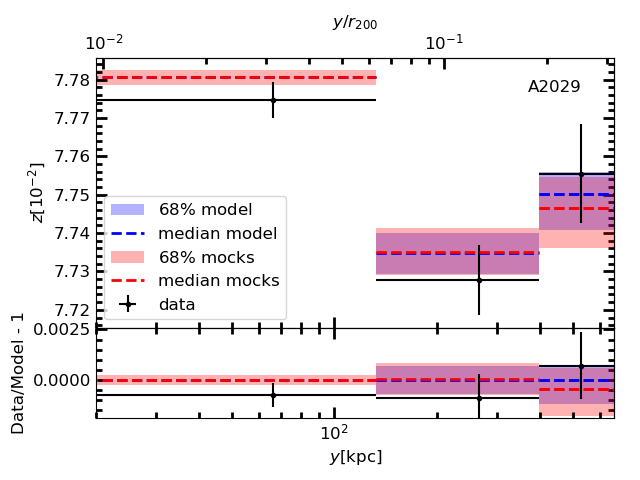}
   \includegraphics[width=0.49\textwidth]{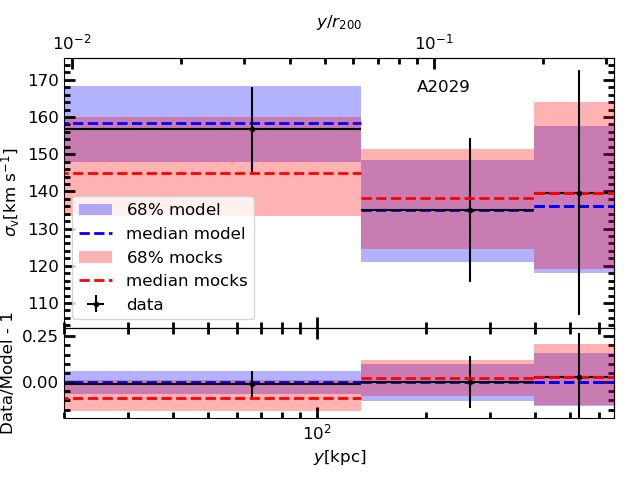}}
   \includegraphics[width=0.49\textwidth]{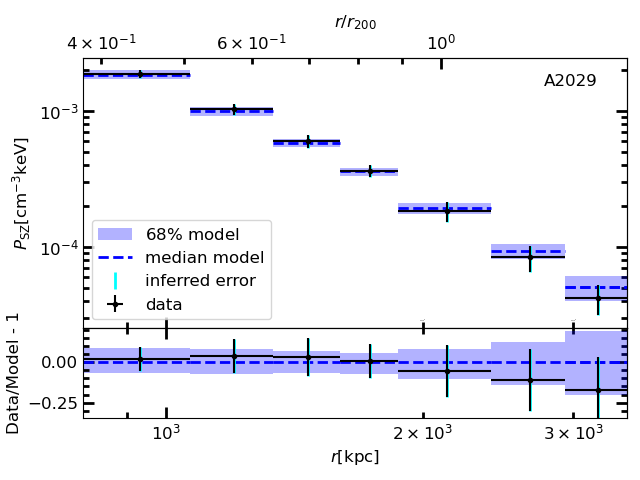}
    \caption{Profiles (upper panel) and residuals (lower panel) of the brightness ($\mathcal N$; left-top panel), temperature ($T$; right-top panel), redshift ($\zu$; left-middle panel), LOS non-thermal velocity dispersion ($\sigma_\mathrm{v}$; right-middle panel) and of SZ-derived thermal pressure ($P_\mathrm{SZ}$; bottom panel). 
    The observational measurements are the black dots with the horizontal and vertical error bars indicating the bin extents and the $1\sigma$ statistical errors, $\sigmastat$, respectively.
    The cyan vertical error bars indicate the total errors, $\sqrt{1 + \epsilon^2} \sigmastat$ (see Sect.~\ref{sec:stat}, for details).
    The corresponding median values computed from the model and the corresponding median measurements from the mock spectra (see Sect. \ref{sec:simulation}, for details) are the blue and red dashed lines, respectively, while the shaded rectangles span vertically the 16th -- 84th  percentile intervals of the mock measurements.
     Both the uncertainty in the parameter values from the posterior and the Poisson noise in the mock spectra contribute to the error of the mock measurements, which is estimated as follows.
    We start with 100 measurements of the quantity $W_m$, extracted from each mock spectrum. 
    We randomly generate 100 values for each mock measurement according to the corresponding error.
    From the distribution consisting of all the values generated for every mock datum at a given bin, we measure the 16th -- 84th percentile interval.
    }
    \label{fig.fitting}
\end{figure*}

Following Sect. 5 of \citetalias{Bartalesi25}, we evaluate the accuracy of our model in reproducing the observational measurements through the residual, defined at each spatial radial bin as the difference between the observational measurement and the corresponding model value, normalized to the latter.
The comparison of the values computed from the model for this representative set of the posterior sampling with those obtained by modeling the signal from X-ray and microwave observations is presented in Fig. \ref{fig.fitting}.
The values of $\Niii$, $\Ti$ and $\PSZj$ in our model reproduce well the corresponding values for all the values of $i$ and $j$, with median residuals of $\approx 0.005$, $-0.001$ and 0.007, respectively.
However, we note that, in the radial range (300 -- 700) kpc, corresponding to $i = \{7, ..., 11 \}$, the values of $\Niii$ and $\Ti$ between the median model and the observational constraints differ by a factor up to 3.2 times the total (statistical plus systematic) uncertainty, probably due to local fluctuations either in the density or temperature profiles, whereas the pressure profile, obtained by deprojecting and combining gas density and temperature, is well modeled within $1\sigma$ uncertainty.
In Sect. \ref{sec:intrinsic} we introduced a three-zone polytropic distribution, with parameters $\gamCEN$ and $\Rcen$ (the transition radius from the central regions, with polytropic index $\gamCEN$, to those intermediate with $\gamIN$) in addition to those used by \citetalias{Bartalesi24} and \citetalias{Bartalesi25}\footnote{The parameter $R_\mathrm{break}$ of \citetalias{Bartalesi24} and \citetalias{Bartalesi25} was renamed as $R_\star$ in this work.} (see their Sects. 3). 
The MCMC run constrains the marginal posterior distribution of $\Rcen$ to a well limited radial range, approximately corresponding to $\approx$ 100 kpc ($\approx$ 0.05 $r_{200}$) from the center, and the marginal posterior distribution of $\gamCEN$ to values more than $10\sigma$ discrepant from the median $\gamIN$ (see Table \ref{tab:prior}).
This implies that the density and temperature measurements in the core of A2029 require two polytropic components transitioning at a radius of $\approx$ 0.05 $r_{200}$. 
We note that, similarly to ours, \citet{Ghirardini_poly} find polytropic index increasing with increasing radius not only for A2029, but also for the other three cool-core clusters in their X-COP sample.
The parameter $\eta$ (see Sect. \ref{sec:from_to}) inferred in our model, $0.96 \pm 0.1$, is consistent within $1\sigma$ credible interval with the median of the $\eta$ distribution for the other X-COP clusters, inferred by \citetalias{Bartalesi25} (see their Appendix A).
In addition, the consistency of $\eta$ with unity, which is the expected value for a spherically-symmetric ICM distribution, implies that current data support the absence of offset between the X-ray and SZ-derived thermal pressure profiles \citep[e.g.,][]{Kozmanyan19}.
The median values of $\epsN$, $\epsT$ and $\epsSZ$ inferred from our model are $0.40$, $0.13$ and $0.06$, respectively (see the corner plot in Appendix \ref{sec:corner}). 
These values indicate that the contribution of the systematic errors to the total error budget is subdominant with respect to the statistical uncertainties.
The median inferred value of $\epsN$ is larger than those of $\epsT$ and $\epsSZ$, but is consistent with the lower end of the distribution of the corresponding parameter reported in Sect.~5.1 of \citetalias{Bartalesi25}.
In all the $m$-th rectangular pointings, the values of $\zum$ and $\sigmavm$ from the intrinsic properties of our model are consistent with the observed ones, with median residuals of 0.001.
In Sect.~\ref{sec:simulation}, generating and analyzing synthetic counterparts of all the XRISM/Resolve observed spectra for the ICM in our model, we derive mock measurements of $\zum$ and $\sigmavm$ that are compared with those inferred here.

From the posterior sampling, we computed a few relevant quantities for the interpretation of the results.
At a given point in the equatorial plane, we evaluate the rotation velocity, $u_\phi(R)$, through eq. (\ref{eq:rotation}), $\sigmaturb(R, 0)$ (defined in Sect. \ref{sec:from_to}) and the 1D velocity dispersion of the plasma, $\sigmagas(R, 0) = \sqrt{P(R, 0) / \left[\mu \proton n(R, 0)\right]}$, where $P(R, 0)$ and $n(R,0)$ are obtained as described in Sect. \ref{sec:intrinsic}.
As a measure of the dynamical importance of rotation, we computed, as a function of radius $R$, the ratio between $u_\phi(R)$ and $\sigmagas(R,0)$ in the equatorial plane, $u_\phi/\sigmagas$.
We compute the turbulent-pressure-to-total-pressure ratio $\alphaturb$, defined by Eq.~(\ref{eq:alpha_turb}), to quantify the dynamical importance of the turbulence.

\begin{figure*}
   \centering
   \hbox{
   \includegraphics[width=0.49\textwidth]{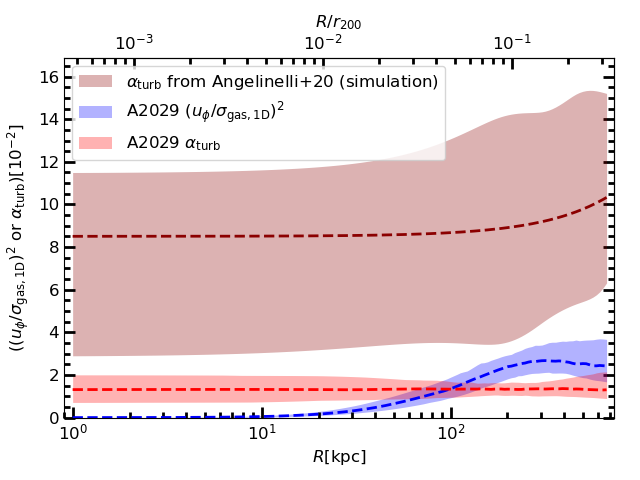}
   \includegraphics[width=0.49\textwidth]{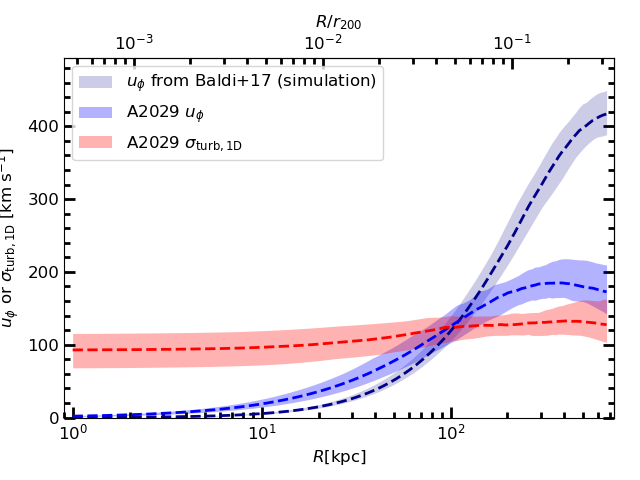}}
    \caption{Squared rotation-velocity-to-velocity-dispersion ratio, turbulent-to-total pressure ratio (left panel), rotation velocity and turbulent velocity dispersion (right panel) as a function of the cylindrical radius. 
    We overplot the median and average profiles of turbulence and rotation measured in simulated clusters by \citetalias{angelinelli20} and \citetalias{BALDI17} (dark red and dark blue regions; see text, for details), respectively. 
    The dashed lines indicate the median profiles; the shaded regions represent the 16th -- 84th percentile interval profiles.
    The \citetalias{angelinelli20} and \citetalias{BALDI17} profiles are obtained as in Figs. 2 of \citetalias{Bartalesi25} and 3 of \citetalias{Bartalesi24}, respectively.
    The 16th -- 84th percentile interval of \citetalias{BALDI17} profile refers to the uncertainty in the posterior distribution of our model parameters. 
    }
    \label{fig.kinematics}
\end{figure*}

Figure~\ref{fig.kinematics} shows $u_\phi$, $u_\phi / \sigmagas$, $\sigmaturb$, and $\alphaturb$ for the representative set of the posterior sampling mentioned above as a function of the cylindrical radius $R$. 
We limit ourselves to $R$ in the range (0 -- 668) kpc, corresponding to the range along the positive $y$-direction sampled by the three XRISM/Resolve pointings.
Both $\sigmaturb$ and $\alphaturb$ exhibit nearly radial constant profiles, with median values of 100 $\kms$ and 0.02, respectively.
Instead, the median profile of both $u_\phi$ and $u_\phi/\sigmagas$ increases outwards, reaching a median peak of 200 $\kms$ and of 0.15 at $R$ between 200 and 600 kpc.
The broad 16th -- 84th percentile interval for $\Rpeak$, spanning (250 -- 540) kpc (see Fig. \ref{fig.corner}), indicates that the current XRISM/Resolve data put weak constraints on the position of the $u_\phi$ peak.
In Fig.~\ref{fig.kinematics}, we also plot the profiles representative of the average properties of clusters formed in cosmological N-body hydrodynamical simulations. 
In particular, we take the average $u_\phi$ profile of a maximally rotating sample selected within the MUSIC-2 sample\footnote{The MUSIC-2 sample \citep{Sembolini13} is composed by 258 clusters with $\Mtwo > 7 \times 10^{14}$ M$_\odot$ re-simulated with a zoom-in technique in a cosmological environment. \citetalias{BALDI17} classified as rotating clusters those with a spin parameter (see their Eq. 2, for its definition) larger than 0.07 that represents 4\% of \cite{Sembolini13} population. We present the average properties of this rotating population in the radiative run, which refers to the simulation including radiative cooling and subgrid models for feedback from active galactic nuclei and from massive stars. However, as discussed in \citetalias{BALDI17}, the average rotational properties of these systems are similar to the non-radiative run.} and the median $\alphaturb$ profile of the Itasca sample from \citetalias{BALDI17} and \citetalias{angelinelli20} \footnote{The Itasca \citep{Vazza17} sample is composed by nine clusters with $0.5 < M_{100} / (10^{14} \mathrm{M}_\odot) < 4$ that were re-simulated with a higher resolution in the cosmological environment. Taking snapshots of these clusters at redshift 0 < $z$ < 2, separated by at least one dynamical time of the cluster, they built a sample of 68 clusters and showed that its mass function closely resembles that predicted at $z=$ 0.}, respectively.
Similarly to our $u_\phi$, the \citetalias{BALDI17} profile increases outwards, but their median peak of $u_\phi \approx$ 400 $\kms$ in the plotted range is inconsistent to a 1$\sigma$ credible interval with our inference. 
Instead, although the shape of our $\alphaturb$ profile seems not too much different from that of \citetalias{angelinelli20}, the latter is slightly increasing outwards and inconsistent to a 1$\sigma$ credible interval with our inferred values.

In \citetalias{XRISM_a2029_2_25}, the same spectral data of XRISM/Resolve used in this work are analyzed, and an effective non-thermal velocity dispersion (obtained by adding in quadrature the measured velocity dispersion and the bulk motion) is combined with the thermal component to construct a total pressure in equilibrium in an NFW potential. Our $\alphaturb$ is consistent within $1.5\sigma$ credible interval with the corresponding median value of the non-thermal-to-total pressure ratio (see their Fig. 6). 
However, our median profile does not support the outward-decreasing trend they report.
The improvements in the underlying physical model and in the accuracy of the modeling of the PSF effect (see Sect.~\ref{sec:PSF}), as demonstrated by the comparison with the results of the spectral simulations in Sect.~\ref{sec:simulation}, may explain why our model has a flatter profile than \citetalias{XRISM_a2029_2_25}.

For the Monte Carlo spectral simulations run in Sect. \ref{sec:setup}, we use the "extflat" mode in the software {\tt HeaSim} that assumes circular symmetry of the distribution of the quantities of our model in the plane of the sky.
Given that the characteristic of the rotation is a dipole-like pattern in the map of the projected LOS rotation velocity, with positive values on one side of the cluster and negative on the opposite side, this assumption appears to be less justified for the rotation observables.
This dipole-like pattern can be approximately modeled by considering Eqs. (\ref{eq:app_u}) and (\ref{eq:app_sigma}) instead of Eqs. (\ref{eq.ulosm}) and (\ref{eq.sigmam}) in the derivation of $\zum$ and $\sigmavm$ in the analysis above.
Using Eqs. (\ref{eq:app_u}) and (\ref{eq:app_sigma}) and assuming the same prior as in Table \ref{tab:prior}, we refit our model to the same observational measurements via a MCMC sampling.
We find that the rotation and turbulence profiles, obtained from the latter inference, are similar to those of Fig. \ref{fig.kinematics}.

\section{Comparison with the results of the spectral simulations}
\label{sec:simulation}
The {\tt Heasim} software, part of the standard {\tt HEASoft} package, is an event generator specifically developed for Monte Carlo spectral simulations with high-energy focusing telescopes, including XRISM.
For a pointing on the sky, it produces a synthetic event file that records the photons generated according to an input X-ray emissivity and registered through an XRISM/Resolve-like instrument, making it an ideal tool for forward-folding the diffuse X-ray emission associated with our dynamical model for A2029.
For the representative set of the posterior sampling extracted in Sect.~\ref{sec:posterior}, we want to produce synthetic counterparts of all three XRISM/Resolve spectra under consideration and to measure the shifts and broadenings of their emission lines.
Section~\ref{sec:setup} details the procedure used to generate and analyze synthetic counterparts of all the XRISM/Resolve event files for the three pointings (see Sect. \ref{sec:resolve}) for the generic parameter vector $\vec \theta$ of our model. 
Sect. \ref{sec:results} presents and discusses the results of the analysis of the synthetic spectra for values of $\vec \theta$ belonging to the representative set of the posterior distribution.

\subsection{Setup of the spectral simulations}
\label{sec:setup}

Assuming that the position of the minimum of the gravitational potential of our model coincides with the sky coordinates of the $m=1$ pointing center, we define a circular region extending out to 11.2 arcmin ($\approx$ 950 kpc at $z_0$) from the center in the plane of the sky. 
This region is divided into 90 circularly-symmetric, concentric annuli, linearly-spaced in radius.
Consistent with Sect. \ref{sec:model}, the $i$-th annulus has inner and outer radii $\Ri$ and $\Rplus$, respectively. 
At any given pointing $m$ and annulus $i$, fixing $\Nh$ to the same value as in Sect. \ref{sec:resolve}, we compute in {\tt XSPEC} the emission model {\tt tbabs}$\times${\tt bapec}$_\mathrm{m,i}$, with the parameters $\Normi$, $\Ti$, $\Zi$, $z_{\mathrm{u}, m, i}$ and $\sigma_{\mathrm{v}, m, i}$ defined for a given $\vec \theta$ as follows.
Note that the quantities without the subscript $m$ do not depend on the characteristics of each pointing.
We take $\Normi = \Normizero$ and $\Ti = \Tizero$, where $\Normizero$ and $\Tizero$ are obtained from Eqs. (\ref{eq.Norm}) and (\ref{eq.Tsl}), respectively. 
$\Zi$ is evaluated at the radius $(\Ri + \Rplus) / 2$ according to Eq. (\ref{eq.mernier}).
We derive $u_\mathrm{i,0}$ and $\sigmavi$ from Eqs. (\ref{eq.uew}) and (\ref{eq.sigmaew}), respectively.
We recall that the rotation manifests itself as a contribution to $\sigmavm$ and $\zum$ in the $m=1$ and $m = \{2, 3\}$ pointings, respectively (see Sect. \ref{sec:stat}, for details).
For $m = 1$ and a given $i$, $\um = 0$ in Eq. (\ref{eq:z_u}) and $\sigma_{\mathrm{v},1,i} = \sqrt{ \sigmavi^2 + u_{i,0}^2}$. 
It is important to note that in this Section the value of $\um $ in Eq. (\ref{eq:z_u}) depends on both the pointing $m$ and the annulus $i$.
For $m = \{2, 3\}$ and a given $i$, $\um = -u_{i,0}$ in Eq. (\ref{eq:z_u}), and $\sigma_{\mathrm{v},m,i} = \sigmavi$.

As standard in the framework for forward-folding the emission from an extended source through the instrumental responses, the vignetting\footnote{The vignetting is the decrease in the measured count rate switching from on-axis to off-axis observations of a point-like source.} and PSF models are treated separately from the ARF.
We take the $ARF_m$ of a point-like source being in the center of the Resolve FoV for the $m$-th pointing from Sect. \ref{sec:resolve}.
It already accounts for the photon loss of a point-like source due to PSF scattering outside the detector.
Also {\tt HeaSim} simulates this effect in a manner appropriate for the extended emission model we provide (see below).
If we use the standard $ARF_m$ for the {\tt HeaSim} simulations, we would model twice the PSF photon scattering.
To avoid this, we use for the {\tt HeaSim} simulations a corrected ARF of a point-like source, defined as $ARF_{\mathrm{fc},m} = ARF_m / 0.804$, where 0.804 is the fraction of the PSF model, for a centered point-like source, lying inside the detector region\footnote{This value is reported in the {\tt xaarfgen} log output when run for every $m$ (see Sect. \ref{sec:resolve}).}.
For the $m$-th mock observation, {\tt Heasim} uses $RMF_m$, $ARF_{\mathrm{fc},m}$ and $NXB_m$\footnote{We recall that this background file has been built while excluding the pixel 27 that, instead, in the simulation is included.} (see Sect. \ref{sec:resolve}), the energy-dependent PSF and vignetting files from the Cycle 1 calibration files\footnote{\url{https://heasarc.gsfc.nasa.gov/docs/xrism/proposals/responses.html}}.
The sky coordinates are fixed to the RA and DEC values recorded in the corresponding $m$-th event file.
To have similar relative error on the inferred values of the redshift in the three pointings, we rescale the exposure time to be set in the {\tt HeaSim} spectral simulation for a given mock observation by the emissivity. 

For the $m$-th pointing and the $i$-th annulus, we generate a photon distribution in position and in energy, $U_{m,i}(Ra, Dec, E)$, via a Monte Carlo sampling, using the "extflat" mode of {\tt HeaSim} that considers the emission model {\tt tbabs}$\times${\tt bapec}$_{m,i}$ to be uniformly distributed over all the $i$-th plane-of-the-sky annulus and to vanish elsewhere. 
Then, the ray-tracing simulations, run with {\tt Heasim}, redistribute the photon distribution $U_m(Ra, Dec, E) \equiv \sum_{i=1}^{90} U_{m,i}$ in position in the focal plane of the detector and in energy according to the Resolve characteristics.
The simulation results in an event file for each value of $m$ containing the counts that a Resolve-like instrument would measure from the polytropic distribution of turbulent and rotating ICM defined by $\vec \theta$. 

Considering all the counts reported in the $m$-th synthetic event file except for those detected in pixel 12, we extract the $m$-th spectrum.
Following the same procedure as for the real $m$-th spectrum (see Sect. \ref{sec:resolve}, for details), we bin and, then, fit the $m$-th mock spectrum to infer $\zum$ and $\sigmavm$ of a uniform and homogeneous plasma in CIE.
We repeat the procedure described above for each pointing for each selected $\vec \theta$.

\subsection{Results of the spectral simulations}
\label{sec:results}
In addition to the metallicity, the density and temperature of the ICM, which mainly shape the spectral continuum in the X-rays, regulate the fluxes of the emission lines, whose shifts and broadenings encode the spectral signatures of rotation and turbulence that we want to constrain with XRISM/Resolve data in Sect. \ref{sec:posterior}.
For a given $\vec \theta$, as the model provides a distribution of the density and temperature, which was constrained in Sect. \ref{sec:posterior} to reproduce the XMM measurements, the model provides, via the generation of the photon distribution $U_m(Ra, Dec, E)$, precise predictions on the count rate as a function of the spectral energy in the $m$-th XRISM/Resolve pointing.
In line with the statistical approach adopted for our study, in this Section we compare the median count rate within the $m$-th synthetic spectra of a representative set of the posterior distribution with the corresponding observed value in each energy bin.
Though our approach allows in principle to reconstruct the emission observed in the three XRISM/Resolve pointings according to the XMM observations, the detailed investigation of the match between the XMM and XRISM/Resolve observations lies beyond the scope of this work.

In addition to $U_m$, the combination of  $RMF_m$, $ARF_{\mathrm{fc},m}$, $NXB_m$ and the model for the PSF (see Sect. \ref{sec:setup}, for details) determines the count rate in the $m$-th mock spectrum.
Now, we report the systematics in these components of the spectral simulation that can introduce offsets between the $m$-th count rates observed by XRISM/Resolve and inferred using the posterior of our model.
The requirement for the precision of in-flight calibration of the XRISM/Resolve effective area is ±10\%, both over a broad energy range, as well as within several narrower energy bins (of a few keVs size) \citep[see ][]{Miller21}.
This work is ongoing, and current discrepancies between XMM and XRISM Resolve and Xtend can be up to $\pm$ 20\%, as reported by \citet{Xrism_calibration}, leading to offsets between the XRISM/Resolve and XMM count rates ranging from 0.8 to 1.2.
We recall that {\tt HeaSim} assumes a PSF model that reproduces the on-ground azimuthally-averaged measurements for the Soft X-ray Imager of the mission Hitomi.
Given that the PSF distribution is expected to scatter out of the central pointing a significant fraction of the photons originating from the central regions of a cool-core cluster with a very peaked density profile, such as A2029, the PSF uncertainties\footnote{For sake of completeness, we report that, based on a comparison of pixel-to-pixel relative count rates and with the corresponding predictions of ray-tracing simulations, \citetalias{XRISM_a2029_2_25} conservatively quantified that the uncertainties in the XRISM/Resolve PSF can introduce an offset within the range 0.7 -- 1.3 (see their Sect. 3.2).} can also realistically produce an offset in the range (0.9 -- 1.1) for $m = 1$ and (0.95 -- 1.05) for $m = \{2, 3\}$.

For the sake of simplicity, in Sect. \ref{sec:setup}, we assumed the X-ray emission to be circularly-symmetric in the plane of the sky.
Under this assumption, the ratio between the total observed and mock count rates in the (3 -- 10) keV energy range, $\Rcount$, is\footnote{The relative error on $\Rcount$ is estimated by adding in quadrature the relative uncertainties of broad-band mock and observed count rates. We notice that the statistical uncertainty of the count rate of each synthetic spectrum has two contributions: the Poisson noise introduced in the spectral simulation and the uncertainty in the parameter values in the posterior.} 1.18 $\pm$ 0.01, 1.17 $\pm$ 0.02 and 1.09 $\pm$ 0.02 for $m=\{1, 2, 3\}$, respectively.
Note that the quoted errors are only statistical, while as argued above, systematic uncertainties at the level of 20-30\% dominate the error budget, but are hard to quantify at this time.
However, the XMM mosaic map analyzed by \citetalias{Ghirardini19} shows enhanced emission in the NE azimuthal sector, the region covered by all XRISM/Resolve pointings.
For each pointing, we adopt a correction factor, $\Raz$, defined in Appendix \ref{sec:NE} as the ratio between the count rate in the NE sector and the azimuthally averaged value, independently of the spectral energy.
After multiplying the mock count rate in each energy bin by $\Raz$, the corrected $\Rcount$ becomes $1.04 \pm 0.02$ and $0.93 \pm 0.02$ for $m = 2$ and $m = 3 $, respectively.
For the central pointing ($m = 1$), the azimuthal variations in the X-ray emission are unlikely to produce an offset in  $R_{\mathrm{count},1}$ as substantial as for $m = \{2, 3\}$.
In conclusion, the corrected $\Rcount$ is consistent with unity, i.e.\ no offset between predicted and observed count rates, within the combined systematic (due to PSF model and effective area measurements) and statistical uncertainties for any $m$.

\begin{figure}
   \centering
   \includegraphics[width=0.49\textwidth]{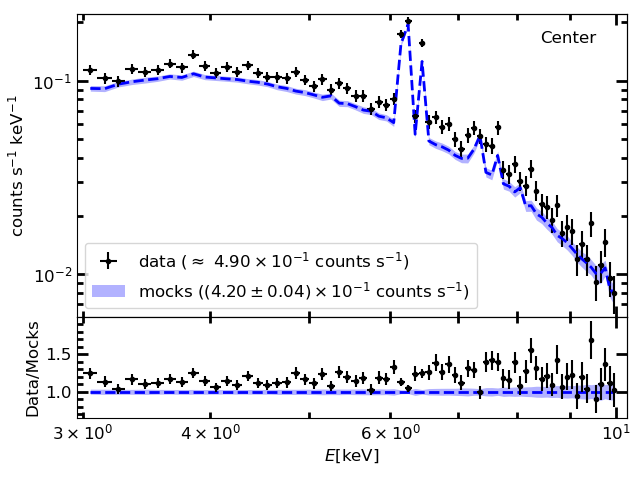}
   \includegraphics[width=0.49\textwidth]{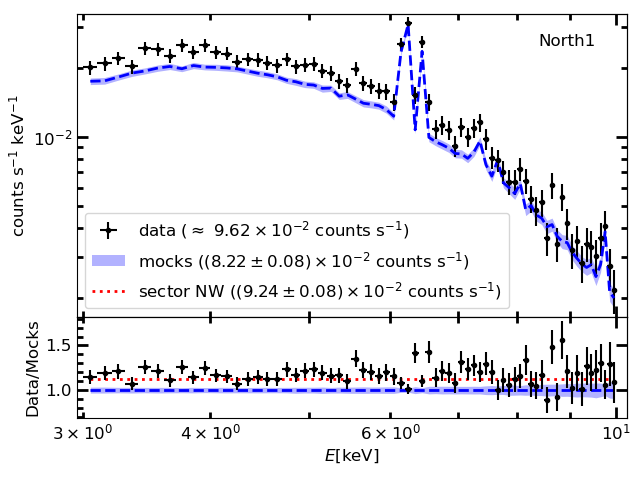}
   \includegraphics[width=0.49\textwidth]{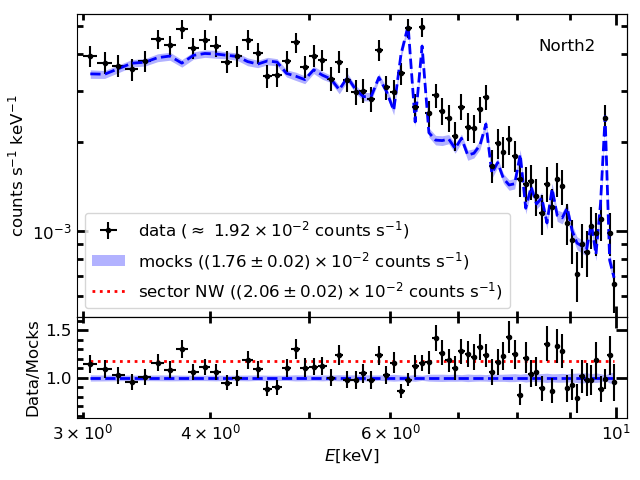}
    \caption{
    Count rates (upper panels) and residuals (lower panels) of the $m=\{1, 2, 3\}$ (top, middle and bottom panels, respectively) XRISM/Resolve pointings of A2029, as a function of the spectral energy.
    The median count rate of the mock spectrum and its 16th--84th percentile interval are represented by the blue dashed line and blue shaded region, respectively.
    The observational data are the black dots, whereas the black vertical error bars indicate the $1\sigma$ uncertainties and the horizontal error bars the bin extent.
    The value of $\Raz$ for $m=\{2, 3\}$ (see text) is indicated by the horizontal red dotted line in the lower panels.
    }
    \label{fig:mocks}
\end{figure}

Figure~\ref{fig:mocks} presents for each pointing the comparison between the mock and observed count rates in spectral energy bins in the (3 -- 10) keV (their total count rates are reported in the legend). 
We find no statistical evidence for either a trend of the ratio between observed and mock count rates with the spectral energy or an offset in energy bins of a few keV.
The best-fit metallicity measured in each mock spectrum is consistent within a 2$\sigma$ confidence interval with the corresponding observed value.

Now that we have established that the density, temperature and metallicity distributions of the plasma in our model are consistent with the X-ray emission resolved by XRISM/Resolve in the three spectra, we compare the line shifts and broadenings between the $m$-th synthetic and observed spectrum. 
The values of $\zum$ and $\sigmavm$, obtained by fitting the emission model for a uniform and homogeneous plasma in CIE to the $m$-th spectral data, quantify the average shift and broadening of the X-ray emission lines in the $m$-th spectrum.
The middle panels of Fig. \ref{fig.fitting} show that the mock measurements of $\zum$ and $\sigmavm$ reproduce the corresponding XRISM/Resolve values, with median residuals of 0.001, respectively, which are consistent with those reported in Sect. \ref{sec:posterior}.
This comparison demonstrates that the profiles of turbulence and rotation shown in Fig. \ref{fig.kinematics} remain accurate in reproducing the observed values of $\zum$ and $\sigmavm$ in the $m$-th XRISM/Resolve spectrum even when we use the more accurate, but computationally more expensive procedure carried out in this Section.

\section{Summary and conclusions}
\label{sec:conclusion}
We have presented an axisymmetric, equilibrium model that reproduces the observed properties of the ICM in the galaxy cluster A2029.
Building on the formalism used by \citetalias{Bartalesi24} and \citetalias{Bartalesi25}, the model describes a composite-polytropic distribution of a rotating and turbulent plasma in equilibrium in the gravitational potential of a spherically-symmetric NFW dark matter halo. 
The profile of rotation velocity and the distribution of turbulent velocity dispersion are described with relatively flexible functional forms that are consistent with the corresponding average radial profiles derived from the analyses of the MUSIC-2 and Itasca samples of clusters formed in cosmological simulations by \citetalias{BALDI17} and \citetalias{angelinelli20}, respectively (see Sect. \ref{sec:intrinsic}).

We reduced and conducted the spectral analysis of the data from the three XRISM/Resolve pointings for A2029, which were acquired during the Performance Verification phase of the mission and sampled its North-East (NE) direction out to $r_{2500}  \approx 669$ kpc.
In addition to the measurements of the LOS non-thermal velocity dispersion ($\sigmav$) and redshift ($\zu$, probing the LOS component of the ICM velocity) in the three XRISM spectra, we considered the measurements of the normalization of the thermal continuum (which scales with the gas density squared), of the spectroscopic plasma temperature, and of the SZ-derived thermal pressure as obtained by \citetalias{Ghirardini19} in the radial bins defined in the XMM-Newton and Planck maps (see Sect. \ref{sec:data}).

In the comparison with observational data, we assume to observe our model edge-on, in particular with the rotation and symmetry axis orthogonal to the NE direction, and fix the systemic redshift of our model to that of BCG ($z_0$).
We fitted the radial metallicity profile of \cite{Mernier17} to the XMM measurements of \citetalias{Ghirardini19} and assumed the best-fit profile over all our analysis.
In the derivation of $\zu$ and $\sigmav$ from the model, we weighted the kinematic observables over this metallicity distribution in the plane of the sky and, assuming realistic circularly-symmetric profiles for the PSFs of XRISM and XMM, we convolved the observable quantities of our model with them.
Under these assumptions, from the intrinsic properties of the plasma in our model, we computed the observable quantities that approximate those recovered from the spectral analysis (see Sect. \ref{sec:from_to}).
We fitted via an MCMC sampling this model simultaneously to all the aforementioned observational measurements, to infer the posterior distribution of the model parameters (see Sect. \ref{sec:joint}).

According to the posterior distribution of our model, under the same assumptions of metallicity and PSF profiles, we generated an ensemble of synthetic spectral counterparts of the three XRISM/Resolve pointings for our model.
After showing that the synthetic spectrum of each pointing is consistent within the systematic uncertainties with the corresponding observed spectrum, we conducted its spectral analysis as for the observed spectrum (see Sect. \ref{sec:simulation}).

The main results of our work are the following.
\begin{itemize}
    \item Our model accurately reproduces the observational measurements of $\zu$ and $\sigmav$ in the three pointings for A2029, as further demonstrated by comparing the measurements of $\zu$ and $\sigmav$ from synthetic and observed XRISM/Resolve spectra (see Sect. \ref{sec:results}).
    \item The inferred turbulent-to-total pressure ratio has a flat profile across all the radial range (0 -- 650) kpc, with a median value of $\approx$ 0.02, and the inferred rotation-velocity-to-velocity-dispersion ratio peaks at a distance on the NE-direction from the center in the range (250 -- 650) kpc, with a median value of 0.15 (see Sect. \ref{sec:posterior}).
    \item Following Sect. 4.3 of \citetalias{Bartalesi24}, we evaluate the ratio between the hydrostatic mass and the true mass of our halo model, which at $R=r_{2500} \approx 669$ kpc is $0.97 \pm 0.01$.
\end{itemize}

Some of the assumptions underlying our analysis, such as rotation as the only bulk motion contributing to the LOS velocity observed in the off-center XRISM pointings and the position angle of the rotation axis, will soon be tested with an approved XRISM/Resolve pointing toward the eastern region of A2029\footnote{As described in Subsect.~\ref{sec:from_to}, we assume a rotation axis in the plane of the sky orthogonal to the A2029 arm sampled with PV pointings and aligned with this forthcoming pointing. 
In the latter, our model predicts a LOS velocity close to zero and a LOS non-thermal velocity dispersion similar to or higher than in the $m=2$ spectrum.
}.
Further constraints on the presence and amount of rotation could be obtained through a full azimuthal coverage of A2029 with XRISM.

The orientation of the rotation axis in the three-dimensional space is difficult to constrain observationally.
In our analysis, we arbitrarily assumed that the rotation axis is orthogonal to the LOS. This choice of the LOS is advantageous not only because it simplifies the computation of the observables (see Eqs.\ \ref{eq.Norm} - \ref{eq.model_press}), but also because it is informative, allowing us (under the assumption of rotation as the only bulk motion) to put a lower limit on the intrinsic rotation speed, for a given LOS velocity. 
For any other LOS the inferred intrinsic rotation speed would tend to be higher: an extreme case would be that of a LOS parallel to the rotation axis, with no rotation-induced line centroid shift, whatever the intrinsic rotation speed.

\citetalias{XRISM_simulations} find that clusters observed by XRISM/Resolve in regions not strongly affected by AGN feedback -- namely, those outside the central few hundred kpc in cool-core systems or throughout non-cool-core clusters -- typically exhibit turbulent-to-total pressure ratios lower than the medians of their counterparts formed in three cosmological simulations (TNG-Cluster, GADGET-X, and GIZMO-Simba). 
Nevertheless, these measured ratios generally remain within the 68\% scatter of simulated values, except for the two outer pointings of A2029.
In those cases, the median turbulent-to-total pressure ratios, which are below 0.01 \citepalias[see Fig. 5 of][]{XRISM_simulations}, have a probability of less than 1\% of occurring within the simulated cluster population (see their Section 4.2 for discussion).
Our inferred rotation and turbulence profiles for the ICM in A2029 lie significantly below those originally derived by \citetalias{BALDI17} and \citetalias{angelinelli20} (see Sect.~\ref{sec:posterior} and Fig.~\ref{fig.kinematics}). Consequently, our complementary analysis supports the \citetalias{XRISM_simulations} finding that the turbulent support in observed clusters tends to fall below that measured in these simulations.
However, our flat profile of the turbulent-to-total velocity dispersion ratio, with a median value above 0.01, reduces the tension between A2029 and its simulated analogs in the outer pointings reported by \citetalias{XRISM_simulations}. 
Our turbulent-to-total pressure ratio places A2029 at the lower end of the simulated cluster distribution, consistent with the behavior of all other objects observed in regions not strongly influenced by AGN feedback.
The accurate forward-modeling of the PSF scattering effect in deriving the model quantities analogous to the measurements may explain differences from the results of \citetalias{XRISM_a2029_2_25}, which were adopted by \citetalias{XRISM_simulations}. 
These discrepancies further highlight the importance of employing more sophisticated model–data comparison methods, such as those presented in this work, potentially including Monte Carlo spectral simulations.

Using optical data, \citet{castellani25} investigated the possible rotation of the galactic component in A2029. 
They detected a statistically significant dipole-like pattern in the redshift distribution of member galaxies between the northern and southern regions of the cluster, with approaching velocities in the north, as for the ICM (see Sect. \ref{sec:resolve}), extending from a few tens of kpc out to beyond $r_{500}$. 
Interpreting this pattern as the rotation of the system of member galaxies, they derived a rotation axis along the NW -- SE direction, broadly consistent with that assumed in our analysis. 
The measured peak rotation velocity and rotation-velocity-to-velocity-dispersion ratio, 220~$\kms$ and 0.18, respectively, agree within the $1\sigma$ credible interval with our inferred values.
The formation history of A2029 likely determines the relationship between the rotational curves of its main baryonic components. 
For example, \citet{Roettiger00} predict a rotating plasma and a static galactic component in an off-axis post-merger cluster. 
If future XRISM/Resolve observations confirm our results, they would indicate a remarkable agreement between the rotational properties of the galactic and plasma components of A2029, offering valuable constraints on its formation scenario.

The validity of hydrostatic mass estimates of galaxy clusters at $\sim$Mpc radii, crucial for cosmological parameter inference, could be robustly assessed through spatially resolved, high-resolution X-ray spectroscopy of cluster outskirts with future missions such as ESA’s NewAthena \citep{newathena} and the Chinese mission HUBS \citep{hubs}.

\begin{acknowledgements}
We thank the referee for the constructive comments that helped improve the paper.
T.B. thanks the financial support from the European Union NextGenerationEU.
S.E. thanks the financial contribution from {\it Theory Grant / Bando INAF per la Ricerca Fondamentale 2024} on ``Constraining the non-thermal pressure in galaxy clusters with high-resolution X-ray spectroscopy'' (1.05.24.05.10).
This work extensively used the Python packages Numpy \citep{Numpy20}, Scipy \citep{scipy}, Matplotlib \citep{matplotlib}, emcee \citep{emcee}, Cython \citep{cython} and Astropy \citep{astropy18} and the HEaSArch softwares Xspec and HeaSim.

\end{acknowledgements}

\bibliography{refs}{}
\bibliographystyle{aa}

\begin{appendix}

    \section{Convolution with the instrumental point spread function and weighting over the metallicity}
    
    \subsection{Point spread function convolution of the observable quantities}
\label{sec:PSF}

To emulate the effect of photon scattering caused by the telescope PSF on the observable quantities derived from spectral analysis, we convolve the spatial distributions of the corresponding quantities of our model in the plane of the sky with a circularly symmetric PSF profile.

To this purpose, we define the function 
\begin{equation}
    \label{eq:func_PSF}
    K(d) =  \left\{
  \begin{array}{ll}
    k(d) & \text{for } {d \leq d_{\max},} \\
    0 & \text{for } {d > d_{\max},}
  \end{array}
\right.
\end{equation}
where $k(d)$ is the PSF profile, $d_{\max}$ a truncation radius and $d$ the distance between two points in the plane of the sky.
When convolved with K, $\Qlos(\Rbar)$ becomes
\begin{equation}
    \label{eq.conv}
    \QK(\Rhat) = \frac{\int_{0}^{\infty} \Rbar d\Rbar \Qlos (\Rbar)  \int_0^{2 \pi} d\phibar  K(||\vec\Rhat - \vec\Rbar||)}{\int_{0}^{\infty} \Rbar d\Rbar \int_0^{2 \pi} d\phibar K(||\vec\Rhat - \vec\Rbar||)}
\end{equation}
where $\vec\Rhat = (\Rhat, \phihat)$ and $\vec\Rbar = (\Rbar, \phibar)$ are two distinct position vectors in the $y$-$z$ plane.

To reduce the computational cost, it can be convenient to limit the evaluation of $||\vec\Rhat - \vec\Rbar||$ to $z=0$ and approximate eq. (\ref{eq.conv}) as 
\begin{equation}
    \label{eq.g_A}
    \QA(\Rhat) = \frac{\int_{0}^{\infty} \Rbar \diff\Rbar \Qlos(\Rbar) K(|\Rhat - \Rbar|)}{\int_{0}^{\infty} \Rbar \diff\Rbar K(|\Rhat - \Rbar|)}.
\end{equation}
We fix $d_\mathrm{max}$ to 1.15 arcmin and 5.75 arcmin for XMM and XRISM, respectively\footnote{For a reference model, we tested values of $d_\mathrm{max}$ increased by 1.1 for both XMM and XRISM PSF profiles, finding insignificant differences from the profiles of Fig. \ref{fig.fitting}.}.
We assume for XMM $k(d) = 1/[1 + (d / \Rcore)^2]^\alpha$, where $\Rcore = 0.145$ arcmin ($\approx$ 12.8 kpc at $z_0$) and $\alpha = 1.68$, as best-fit King-profile values to the \cite{Read11} PSF measurements obtained by combining the three \emph{XMM-Newton} cameras with weight of 0.25 for MOS and 0.5 for pn as in \citet{Croston06}.
For  XRISM/Resolve, we construct $k$ by interpolating with a cubic spline the PSF measurements reported in the {\tt HeaSim} support files\footnote{These PSF data represent azimuthally-averaged values taken from the on-ground measurements for the Soft X-ray Imager on board of Hitomi mission, the twin predecessor of the XTEND imager onboard the XRISM mission. Given the similarity in the instrumentation design of XRISM/XTEND and XRISM/Resolve, their PSFs are expected to be nearly identical.}.
Given that in the comparison with the model we take only the measurements of $\zu$ and $\sigmav$ from the results of the XRISM/Resolve spectral analysis (see Sect. \ref{sec:stat}), we use the PSF data only for 8 keV photons, the energy closest to the Fe-K complex (6 -- 7 keV), which dominates the spectral features used to derive those measurements.

\subsection{Metallicity profile}
\label{sec:metals}
\begin{figure}
   \centering
   \includegraphics[width=0.49\textwidth]{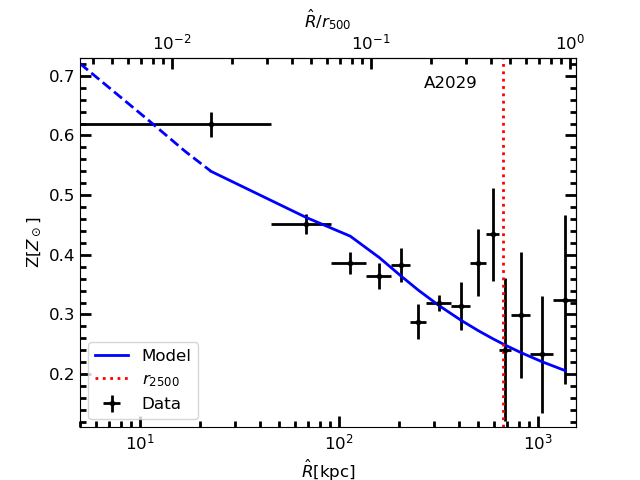}
    \caption{Profile of the metallicity (blue line) as a function of the projected radius $\Rhat$ (the profile in the extrapolation range is dashed).
    The \citet{Ghirardini19} measurements are the black dots, with vertical and horizonthal error bars corresponding to the statistical uncertainties in the spectral analyses and to the extent of the radial bins.
    The XRISM/Resolve pointings extend from the center out to the vertical line.
   }
    \label{fig:metals}
\end{figure}

We assume that the projected distribution of the metals of A2029 in the plane of the sky is circularly symmetric, with profile \citep{Mernier17, Stofanova25}
\begin{equation}
    Z(\Rhat) = \frac{A}{[1 + \Rhat / (r_{500} B)]^C} \left\{1-D\exp \left[-\frac{\Rhat}{r_{500} F} \left(1+\frac{\Rhat}{r_{500} E}\right) \right] \right\} + G,
    \label{eq.mernier}
\end{equation}
where $r_{500} = 1448$ kpc\footnote{$\rfive \approx 1488$ kpc is the value obtained using $\rfive = \left[3 M_{500} / (2000\pi G \rhocrit) \right]^{1/3}$ from the median hydrostatic $M_{500}$ reported in Table 1 of \cite{Ettori19}.}, and
$A$, $B$, $C$, $D$, $E$, $F$ and $G$ are free parameters. 
We evaluate eq. (\ref{eq.mernier}) at the radius $(\Rplus + \Ri) / 2$ for the generic $i$-th circular annulus, with inner and outer radii $\Ri$ and $\Rplus$.
Maximizing a Gaussian log-likelihood, we fit it to the \citetalias{Ghirardini19} measurements.
Fig. \ref{fig:metals} shows the inferred profile, with best-fit parameters  $A = 1.34$, $B = 5.15 \times 10^{-3}$, $C = 4.81\times 10^{-1}$, $D = 4.14\times 10^{-1}$, $E = 1.69\times 10^{-1}$, $F = 2.10\times 10^{-2}$ and $G = 9.68\times 10^{-2}$, compared with the \citetalias{Ghirardini19} measurements.
Throughout this work, we assume that the metallicity distribution in the plane of the sky follows eq. (\ref{eq.mernier}) with these best-fit parameter values.

When weighted over $Z(\Rhat)$ and convolved with $K(d)$ via eq. (\ref{eq.g_A}) (see sect. \ref{sec:PSF}, for details), $\Qlos(\Rbar)$ becomes
\begin{equation}
    \label{eq.f_metal}
    \QZ(\Rhat) = \frac{\int_{0}^{\infty} \Rbar \diff\Rbar Z(\Rbar) \Qlos(\Rbar) K(|\Rhat - \Rbar|)}{\int_{0}^{\infty} \Rbar \diff\Rbar Z(\Rbar) K(|\Rhat - \Rbar|)}. 
\end{equation}
    
    \subsection{1D convolution of rotational properties}
    
    The 1D convolution of the generic projected quantity $\Qlos$ (defined in Eq. (\ref{eq.ffff})), weighted over the projected metallicity profile $Z$ (taken from Eq. (\ref{eq.f_metal})), is
    \begin{equation}
        \label{eq:conv_1D}
        \QL(y) = \frac{\int_{-\infty}^{\infty} d\bar y \Qlos(\bar y) Z(\bar y) K (|y - \bar y|)}{\int_{-\infty}^{\infty} d\bar y Z(\bar y) K (|y - \bar y|)}
    \end{equation}
    where $K$ is defined in Eq. (\ref{eq:func_PSF}).
    The 1D integration in the segment of length equal to the extent of the FoV of XRISM/Resolve for $y\geq 0$ is
    \begin{equation}
        \label{eq:}
        \Lm[Q] = \int_{\ylow}^{\yup} dy \QL(y)
    \end{equation}
    Here Eqs. (\ref{eq.ulosm}) and (\ref{eq.sigmam}) become
    \begin{equation}
        \label{eq:app_u}
        \uewm = \Lm[\nH \nee u_\phi \bar y/R] / \Lm[\nH \nee]
    \end{equation}
     and
    \begin{equation}
        \label{eq:app_sigma}
        \sigmavm = \sqrt{\frac{\Sm\left[\nH\nee \sigmaturb^2\right]}{\Sm[\nH\nee]} + \frac{\Lm\left[\nH\nee \left(u_\phi \bar y / R - u_{i,0}\right)^2\right]}{\Lm[\nH\nee]}},
    \end{equation}
    respectively.
    Here, $\Sm$ is the operator defined in Eq. (\ref{eq:SQm}).
    
    \section{Underestimation of the spectral flux from a sector analysis}
    \label{sec:NE}
    We now quantify the enhancement in the surface-brightness value in the NE direction, i.e.\ the direction of $m=\{2,3 \}$ pointings.
    At any projected radius, $\Rhat$, we consider a value for the azimuthally-averaged surface-brightness, $\SA(\Rhat)$, and a value extracted from a NE sector, $\SN(\Rhat)$. 
    Using Eq. (\ref{eq:SQm}) with $\QZ (\Rhat, 0) = \SN(\Rhat) / \SA(\Rhat)$, we compute this average enhancement in the emission region corresponding to $n$-th Resolve pointing, $\Pn$.
    For sake of simplicity, to estimate the PSF photon scattering, we adopt the ARF matrix from Table 3 of \citetalias{XRISM_a2029_2_25}, denoted as $ARF_{m,n}$, where the index $n$ indicates the emission region and $m$ the observed spectrum.
    Thus, the correction factor is $\Raz = ARF_{m,n} \Pn$.

    \section{Corner plot}
    \label{sec:corner}
\begin{figure*}
   \centering
   \includegraphics[width=0.99\textwidth]{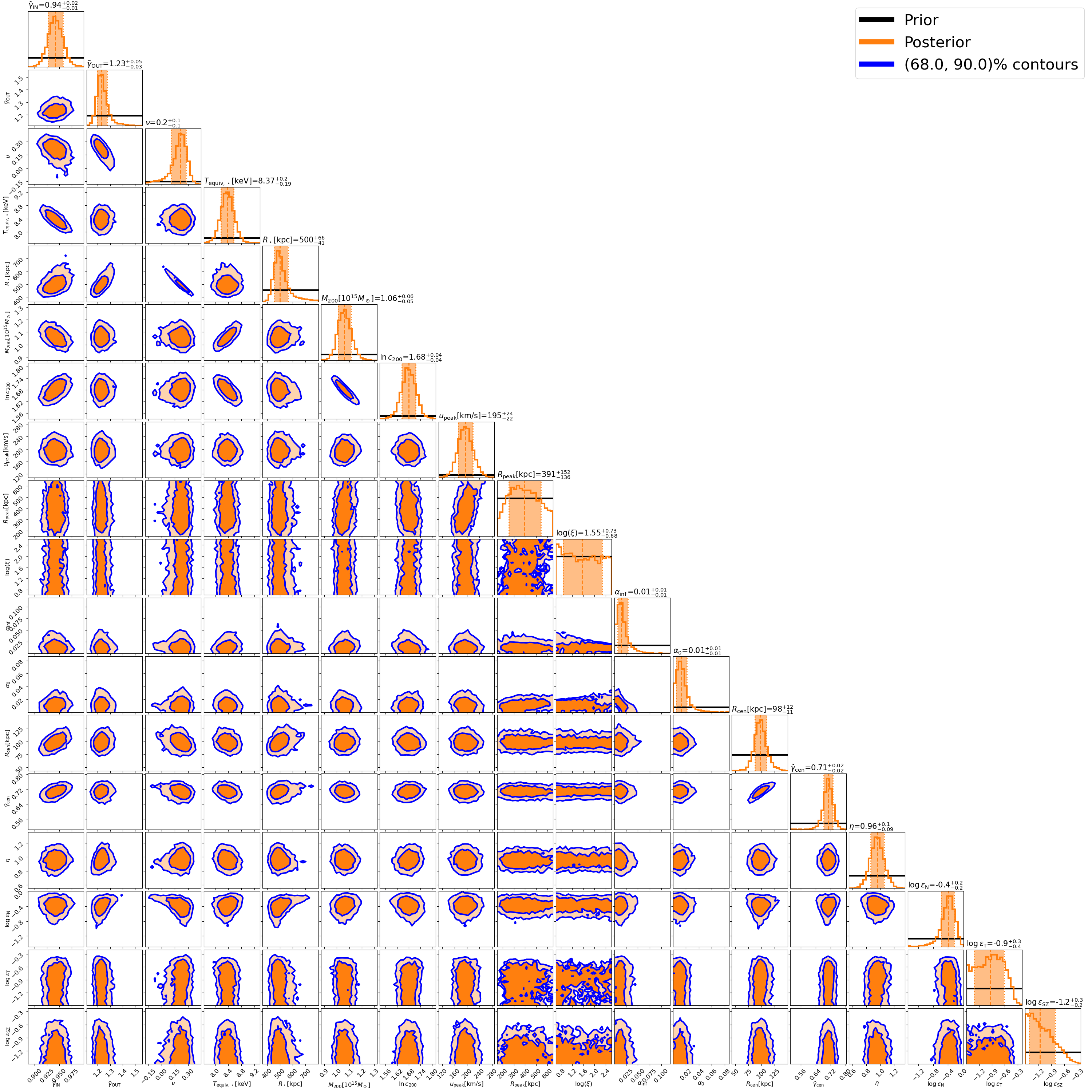}
    \caption{
    Marginal (diagonal panels) and two-parameters joint (off-diagonal panels) distributions of all the free parameters in the MCMC run ($\gamIN$, $\gamOUT$, $\nu$, $\Tequivstar$, $\Rstar$, $\Mtwo$, $\ln c_{200}$, $\upeak$, $\Rpeak$, $\log \xi$, $\alphainf$, $\alphazero$, $\Rcen$, $\gamCEN$, $\eta$, $\log \epsN$, $\log \epsT$ and $\log \epsSZ$; see Sect. \ref{sec:stat}, for details). 
    In each diagonal panel, the black curve and the orange histogram are the marginal prior and posterior, respectively;
    the light-orange vertical band is the 16th-84th percentile interval of the marginal posterior.
    In each off-diagonal panel, the dark-orange and light-orange regions, enclosed by the blue lines, define the 68\% and 90\% credible regions of the joint posteriors, respectively.
    }
    \label{fig.corner}
\end{figure*}

    Fig. \ref{fig.corner} is the corner plot of the posterior distribution analyzed in Sect. \ref{sec:posterior}.

\end{appendix}

\end{document}